\documentclass[structabstract]{aa}
\usepackage{graphicx}
\usepackage{txfonts}
\usepackage{rotating}
\usepackage{lscape}
\usepackage{longtable}
\usepackage[flushleft]{threeparttable}
\usepackage{natbib}
\bibpunct{(}{)}{;}{a}{}{,}
\begin{document}

    \title{The historical record of massive star formation in Cygnus\thanks{Based on observations collected at the Centro Astron\'omico Hispano Alem\'an (CAHA) at Calar Alto, operated jointly by the Junta de Andaluc\'\i a and the Instituto de Astrof\'\i sica de Andaluc\'\i a (CSIC)
    }
   }

   \author{F. Comer\'on\inst{1}
          \and
          A.A. Djupvik\inst{2}
          \and
          N. Schneider\inst{3}
          \and
          A. Pasquali\inst{4}
          }

   \institute{European Southern Observatory, Karl-Schwarzschild-Strasse 2, D-85748 Garching bei M\"unchen, Germany\\
              \email{fcomeron@eso.org}
                     \and
               Nordic Optical Telescope, Aarhus Universitet, Rambla Jos\'e Ana Fern\'andez P\'erez 7, E-38711 Bre\~na Baja, Spain
            \and
           I. Physik Institut, University of Cologne, D-50937 Cologne, Germany
           \and
           Astronomisches Rechen-Institut, Zentrum f\"ur Astronomie der Universit\"at Heidelberg, M\"onchhofstr. 12-14, D-69120 Heidelberg, Germany
 }

   \date{Received; accepted}


  \abstract
     {The Cygnus region, which dominates the local spiral arm of the Galaxy, is one of the nearest complexes of massive star formation, extending over several hundred parsecs. Its massive stellar content, regions of ongoing star formation, and molecular gas have been studied in detail at virtually all wavelengths. However, little is known of the history of the region beyond the past 10~Myr.}
   {We use the correlations between age, mass and luminosity of red supergiants to explore the history of star formation in Cygnus previous to the formation of the present-day associations. The brightness and spectroscopic characteristics of red supergiants make it easy to identify them and build up a virtually complete sample of such stars at the distance of the Cygnus region, thus providing a record of massive star formation extending several tens of Myr into the past, a period inaccessible through the O and early B stars observable at present.}
   {We have made a selection based on the 2MASS colors of a sample of bright, red stars in an area of 84 square degrees covering the whole present extension of the Cygnus association in the Local Arm. We have obtained spectroscopy in the red visible range allowing an accurate, homogeneous spectral classification as well as a reliable separation between supergiants and other cool stars. Our data are complemented with Gaia Data Release~2 astrometric data.}
   {We have identified 29 red supergiants in the area, 17 of which had not been previously classified as supergiants. Twenty-four of the 29 most likely belong to the Cygnus region and four of the remaining to the Perseus arm. We have used their derived luminosities and masses to infer the star formation history of the region. Intense massive star formation activity is found to have started approximately 15~Myr ago, and we find evidence for two other episodes, one taking place between 20 and 30~Myr ago and another one having ended approximately 40~Myr ago. There are small but significant differences between the kinematic properties of red supergiants younger or older then 20~Myr, hinting that stars of the older group were formed outside the precursor of the present Cygnus complex, possibly in the Sagittarius-Carina arm.}
   {}


   \maketitle
%

\section{Introduction}
The Milky Way in the direction of Cygnus harbors one of the nearest giant molecular cloud complexes, containing OB associations, massive star forming regions, recent supernova remnants, interstellar bubbles blown by the combined energetic action of massive stars, and an abundant sampler of virtually all the stages of stellar evolution near the upper end of the initial mass function \citep{Odenwald89,Odenwald93,Bochkarev85,Reipurth08}. The Cygnus region is the  name generally used to refer to this physically coherent complex, a definition that excludes both foreground and background structures physically unrelated to the complex, such as the portions of the Cygnus Rift or the Perseus arm that are also projected in the general direction of the constellation.

The recent massive star forming activity of the region has been thoroughly studied through the O- and early B-type stars of its OB associations, particularly in the rich association Cygnus~OB2 and its surroundings \citep[e.g.,][]{Hanson03,Drew08,Negueruela08,Comeron12, Wright15, Berlanas18,Berlanas19} and the Cygnus~X complex of molecular and ionized gas \citep{Wendker91,Schneider06,Schneider16}. While those stars provide a good picture of the most recent star forming activity, the fast decrease in luminosity with decreasing mass, combined with the considerable extinction toward many areas of the region at visible wavelengths and the general crowdedness at its low galactic latitude, make the stellar census of the region severely incomplete for main sequence spectral types beyond O. Furthermore, the short lifetimes of such massive stars limit their usefulness as probes of the star formation history of the host giant molecular complex to the last $\sim 10$~Myr.

Open clusters are accessible tracers of past star formation over much longer periods. However, many if not most of the stellar aggregates that appear as clusters at early ages become unbound after removal of their parental gas, their members disperse also on timescales of a few Myr, and the identification of clusters in very crowded regions like Cygnus is difficult. For that reason, whereas the presence of a few clusters with ages of tens of Myr in Cygnus shows that star forming activity in the region has taken place over such time spans \citep{Costado17}, the comparison of their number with that of younger aggregates does not provide a suitable representation of the star formation history.

Red supergiants of spectral types K and M provide an alternative approach to probe the history of massive star formation over longer timescales. They are the descendants of stars with masses above $\sim 7$~M$_\odot$ undergoing the helium-burning phase at their cores, during which their radii grow to several AU, their photospheric temperatures drop to near or below 4,000~K, and their bolometric luminosities reach values comparable to those of the brightest O stars \citep{Levesque10,Ekstroem13}. Their temperatures make their spectral energy distributions peak in the near infrared, where they are among the most luminous stars, making their detection easy from large distances and mitigating the effects of interstellar extinction when observing at those wavelengths. The luminosity, as well as the beginning and the duration of the red supergiant phase, are primarily a function of the initial mass of the star, with rotation of the precursor playing a secondary role \citep{Ekstroem12}. For a star of 7~M$_\odot$ with an initial rotational velocity of 40\% of the critical velocity (at which the centrifugal force counterbalances the gravitational acceleration), the red supergiant phase starts at $\sim 50$~Myr and lasts almost 2~Myr, whereas a star with an initial mass of 20~M$_\odot$ rotating at the same fraction of the critical velocity becomes a red supergiant at 9.7~Myr, remaining in that phase for 0.3~Myr. Therefore, red supergiants are bright signposts of past star formation potentially covering the entire lifetime of a giant molecular complex, appearing in that phase for a limited but not negligible time span, which makes them moderately abundant. However, the study of the past massive star formation record of the entire Cygnus region through its red supergiant content is hampered by the lack of a complete, homogeneous census of such stars.

In this paper we present the results of our study of the red supergiant content in the entire Cygnus region. We present a sample of red supergiant candidates selected in an area of 84 square degrees on the basis of their near-infrared colors, which we estimate to be complete down to initial masses $M = 10$~M$_\odot$, with still a substantial completeness fraction down to $M = 8$~M$_\odot$. Visible spectroscopy in the red of all the candidates is presented in order to discard evolved, less massive stars having the same photometric and very similar spectroscopic characteristics. The remaining set, refined with the use of Gaia DR2 data, constitutes an essentially complete sample of red supergiants in Cygnus. We use this sample to investigate the history of massive star formation in the region, and to obtain a crude estimate of its past content.

\section{Target selection\label{target}}

\subsection{Red supergiants in Cygnus\label{rsg}}

The Cygnus complex, which is one of the main structures of the Local Arm of our Galaxy \citep{Xu13}, extends over a vast region covering the interval $71^\circ < l < 85^\circ$, $-2^\circ < b < +4^\circ$, which encompasses the associations Cygnus~OB1, OB2, OB3, OB8, and OB9. This excludes the doubtful, and possibly foreground association Cygnus~OB4 \citep{deZeeuw99}; the also doubtful associations Cygnus~OB5 and OB6 \citep{Uyaniker01}; and the foreground association Cygnus~OB7 \citep{deZeeuw99}.

Red supergiants at distances comparable to those of the OB associations are expected to appear very bright at near-infrared wavelengths, and also have very red colors due both to their low photospheric temperatures and to the foreground extinction, mainly associated with the Cygnus Rift \citep{Straizys15}. We made an initial selection of all the stars in the 2MASS Point Source Catalog in the area defined above having $K_S < 4.0$, $J-K_S > 1.1$. Using the $T_{\rm eff}$ versus intrinsic color in the Johnson-Cousins system from \citet{Kucinskas05}, and transforming to the 2MASS filter system using the transformations from \citet{Carpenter01}, we obtain that the lower limit in $J-K_S$ color selects the supergiants near 4,000~K (having an intrinsic color $(J-K_S)_0 = 0.88$) if reddened by $A_V > 0.7$, where we adopt the extinction law of \citet{Cardelli89} with total-to-selective extinction ratio $R = 3.1$. Cooler stars meet the color selection criteria with even lower values of the extinction. The location of the region beyond the Cygnus Rift implies a moderate level of foreground extinction across the whole region of interest, as confirmed by \citet{Comeron12} and \citet{Berlanas18} who find almost no stars with $A_V < 2$ in Cygnus OB2 and an extensive area around it. Most members of Cygnus~OB1 and OB9 have $A_V > 3$, and those of Cygnus OB3, which is the least obscured of the associations, have $A_V > 1$ \citep{Garmany92}. The chosen limit $J-K_S > 1.1$ is therefore a conservative one expected to be met by all the red supergiants in the region.

To estimate the degree of completeness we assume that all the red supergiants in our sample are in the He-burning stage previous to the blue loop, and that their precursors rotated with an initial velocity of 60\% of the critical velocity. Using the evolutionary tracks of \citet{Ekstroem12}, a red supergiant descending from a star with an initial mass 8~M$_\odot$ would have a bolometric magnitude $M_{\rm bol} = -4.9$ at this stage, which translates into an absolute magnitude $M_K = -7.5$ using the $K-$band bolometric correction scale for red supergiants from \citet{Levesque07}. At the distance modulus $DM = 10.9$ that we adopt for the Cygnus region (see Section~\ref{distance} below) such a red supergiant would be above our detection limit if obscured by $A_V < 5.8$. The distribution of values of the foreground extinction in Cygnus OB2 and surroundings derived by \citet{Comeron12} and \citet{Berlanas18} peaks at $A_V = 4-6$~mag, and is generally lower in the rest of the region. A similar value, $A_V \simeq 5$, is found by \citet{Schneider07} in the neighborhood of the HII region S106 in Cygnus~X, including parts of Cygnus~OB1. We therefore expect most red supergiants with that initial mass to be included in our sample. Foreground extinction can reach higher local values in some specific areas in Cygnus, but those are generally associated with sites of ongoing star formation such as the lines of sight toward the highest column density areas of the Cygnus~X molecular clouds \citep{Schneider06}, often associated with thermal HII regions \citep{Downes66}. We do not expect the presence of evolved objects such as red supergiants in such areas.

Completeness increases rapidly with increasing mass, and for a mass $M = 10$~M$_\odot$ our limit $K < 4.0$ includes all the red supergiants with $A_V < 12$, sufficient to cover the whole extinction depth in the region. Rotation has a small effect on the luminosity at the He-burning stage for red supergiants in the $8-10$~M$_\odot$ range, differing by $\Delta \log L \simeq 0.03$ in the nonrotating case with respect to the rotating case adopted in the sense of nonrotating supergiants being fainter. This translates into $A_V = 5.1$ as the greatest extinction for which a nonrotating red supergiant with initial mass $M = 8$~M$_\odot$ would be included in our sample, and $A_V=11.3$~mag for $M = 10$~M$_\odot$. The latter is still well above the bulk of extinction values found in Cygnus, therefore making our conclusions on completeness only weakly dependent on the assumed initial rotation velocity.

The infrared color-magnitude selection criterion that we use may be expected to include in the sample of candidate supergiants a high, possibly dominating fraction of evolved red giant branch and asymptotic giant branch stars, descending from lower-mass stars with large ages. Many of these have been spectroscopically observed by previous works and have a spectral classification placing them at spectral types M6 or later, at a range of cool temperatures unreachable even by the coolest red supergiants, whereas others have been recognized as long-period Mira variables from their light curves. We have removed both types of stars from our sample, as they are unrelated to the young population of Cygnus.

\subsection{The distance to the Cygnus region\label{distance}}

The distances to associations, clusters and other structures in Cygnus have been traditionally difficult to determine, due to a combination of reasons that include differential extinction and possible extinction anomalies \citep{Terranegra94}, the difficulties in estimating distances based on the observed spectral types and magnitudes of O stars \citep{Mahy15}, the shallow dependency of radial velocities on distance in the directions of Cygnus foreground to the Perseus arm, or contamination of the membership of open clusters due to field crowdedness \citep{Wang00,Costado17}. Furthermore, the distinction among the OB associations included in the area under consideration is mostly historical rather than physical. Some of them, like Cygnus OB1, OB8 and OB9 have been proposed to be part of a single structure \citep{Melnik95}, and it has been suggested that the differences between Cygnus OB2 and OB9 are due to the progression of star formation from lower to higher galactic longitudes rather than to distinct, independent episodes of star formation \citep{Comeron08,Comeron12}. Despite large uncertainties and conflicting results, sometimes yielding discrepant distances by factors up to 2 for the same cluster \citep{Straizys14}, most studies yield values in the 1.3-1.8~kpc range for the associations in the area under consideration \citep[e.g.,]{Garmany92,Hanson03,Kharchenko05,Comeron12,Straizys14,Straizys19,Sitnik15}. Distances to structures in the interestellar medium have been established through evidence of physical links with OB associations, as illustrated by \citet{Schneider07}.

Many of the difficulties encountered by more traditional methods have been overcome in recent years through the determination of trigonometric parallaxes to masers using Very Large Baseline Interferometry (VLBI). This has led to the determination of precise distances to star forming regions in the Cygnus~X molecular complex \citep{Rygl12}, which is closely linked to Cygnus~OB2 \citep{Schneider06,Schneider16}, as well as to other star forming regions outside it \citep{Xu13,Nagayama15}. These latter determinations yield distances around 1.4~kpc, broadly consistent with most determinations based on the stellar component of clusters and associations.

The trigonometric parallaxes measured by Gaia appear nevertheless to raise some tension with those previous determinations. In their analysis of the stellar population of Cygnus~OB2 \citet{Berlanas19} report evidence for a bimodal distribution of stellar distances, with two groups separated by $\sim 400$~pc along the line of sight, but without any obvious angular separation on the sky. A distance of $\sim 1,760$~pc ($DM = 11.15$) is proposed for the main group which, taken at face value, would call into question the physical link, supported by many observations, between most of Cygnus~OB2 and Cygnus~X if the VLBI distances to its star forming regions are adopted for the latter. A second, less numerous foreground population appears at $\sim 1350$~pc, closer to the value adopted in the works cited in the previous paragraph. Distances to several open clusters in the Cygnus region have been presented by \citet{Cantat18,Cantat20}, also based on Gaia~DR2 data as well. The derived distances for most of them are also around 1,700~pc, including the well-studied clusters NGC~6910 and NGC~6913, respectively belonging to the associations Cygnus OB9 and Cygnus OB1. However, in the range of distances relevant to the Cygnus region the effect of the known zero point offset in Gaia~DR2 parallaxes begins to be noticeable. The distance estimates to clusters provided by \citet{Cantat18} are corrected for a systematic offset of 0.029~mas following \citet{Lindegren18}, but \citet{Schoenrich19} have argued that the average offset can be as much as 0.054~mas. Furthermore, the systematic offset may reach up to 0.1~mas locally \citep{Lindegren18,Luri18}. If the Gaia DR2 parallaxes were systematically offset by $0.1$~mas, distances to individual stars and clusters in Cygnus (including the distance to the bulk of Cygnus~OB2 as determined by \citet{Berlanas19}, which is uncorrected for the zero point offset) could be reconciled with the more nearby VLBI distances and with other pre-Gaia determinations that favor the shorter distance. Our sample of red supergiants is of little help to place the distance to the Cygnus region on a firmer standing, as not only are their Gaia~DR2 parallaxes affected like other stars by the same uncertainties and systematic effects, but their very red colors may introduce additional systematics \citep{Drimmel19}. Besides this, red supergiants present specific astrometric difficulties due to their large diameters and large, evolving convective cells \citep{Lopez18}, which can displace their photocenters by angular distances greater than their parallaxes, therefore introducing a virtually random astrometric noise.

Given the uncertainties outlined above, we have given preference in this study to the distances derived through VLBI and to those derived through other methods based on the stellar component. We therefore adopt a distance modulus $DM = 10.9$~mag, corresponding to a distance of 1,510~pc, for all the red supergiants that we assign to the Cygnus region (see Section~\ref{observations}). The corresponding projected area covered in our target selection is therefore $370 \times 160$~pc$^{-2}$. By adopting a single distance we neglect both depth effects along the line of sight, which are expected to be significant especially for the older stars in our sample (see Section~\ref{kinematics}), as well as possible distance gradients across the plane of the sky. Regarding the latter we note that there is no evidence for trends in the parallax as a function of galactic longitude among the nine clusters in the catalog of \citet{Cantat18} that we assign to the Cygnus region, which suggests that its associated structures are spatially arranged in a direction that runs roughly perpendicular to our line of sight.

\section{Observations and spectral classification \label{observations}}

The brightness of red supergiants in Cygnus and the range of interstellar extinction in the area make them generally accessible to spectroscopy at red and far-red visible wavelengths using short exposure times with 2m-class telescopes. This spectral region offers abundant atomic and molecular features sensitive to temperature and surface gravity, enabling the establishment of precise classification criteria in both spectral subtype and luminosity class \citep{Torres93}.

We observed all the stars fulfilling the location, magnitude, and color criteria described in Section~\ref{target}, with the exceptions of those previously classified as M6 or later or as Mira variables in the literature as noted there. This amounts to a sample of 78 stars, which includes all the stars for which no spectral classification has been published thus far, as well as those that have been previously classified in other works as M5 or earlier regardless of the published luminosity class. Indeed, a literature search shows that published classifications are based on differing sets of criteria, having been obtained with a variety of instruments, spectral resolutions, wavelength coverage, and signal-to-noise ratio. We found a number of cases in which significantly different spectral types are assigned to the same star in different works. Furthermore, given the subtle spectroscopic criteria that distinguish young red supergiants from older red giant branch stars, we preferred to observe all the stars to ensure that our sample of red supergiants is based on the uniform application of classification criteria to spectra obtained with the same instrumental setup.

Our observations took place on two nights from August 2-4, 2016, using CAFOS, the facility imager and low-resolution spectrograph at the 2.2m telescope on Calar Alto Observatory in Southern Spain. A grism was used covering the wavelength range $6000 < \lambda (\AA ) < 9000$ providing a resolution $\lambda / \Delta \lambda \simeq 1300$ with a slit $1''5$ wide. Exposure times per target ranged from 5 to 1200 seconds, depending on the $R$ magnitude estimated from the cataloged $JHK_S$ magnitudes. We also observed with the same setup a grid of MK standard stars in the K2-M4 spectral range, with luminosity classes Ia to III, from the catalog of \citet{Garcia89}. The spectrophotometric standard BD$+25^\circ4655$ \citep{Oke90} was used for an approximate relative flux calibration, and a HeNeAr arc lamp was used for wavelength calibration. The extracted, wavelength- and flux-calibrated  spectra of both the Cygnus targets and the reference standards were then normalized to a pseudo-continuum by ratioing them to the best-fitting 7th-degree polynomials in order to remove the effects of the different amounts of foreground extinction toward each object. Spectral classification was carried out using the normalized spectra.

Spectral subtypes were determined by carefully comparing the strength of the TiO and VO features to those of the grid of standards. With only one exception discussed below, all our targets have spectral types in the K-M range. At the resolution of our spectra in the wavelength range covered differences could be discerned between types K2, K5, K7 and M0, but no finer distinction could be made among the spectral subtypes in the K2-M0 range. The greater sensitivity of spectral type with temperature for spectral types later than M0 allowed us to perform a more detailed subtype classification in the M range, with an estimated accuracy of $\pm 0.5$ spectral subclasses.

\begin{figure}[ht]
\begin{center}
\hspace{-0.5cm}
\includegraphics [width=9cm, angle={0}]{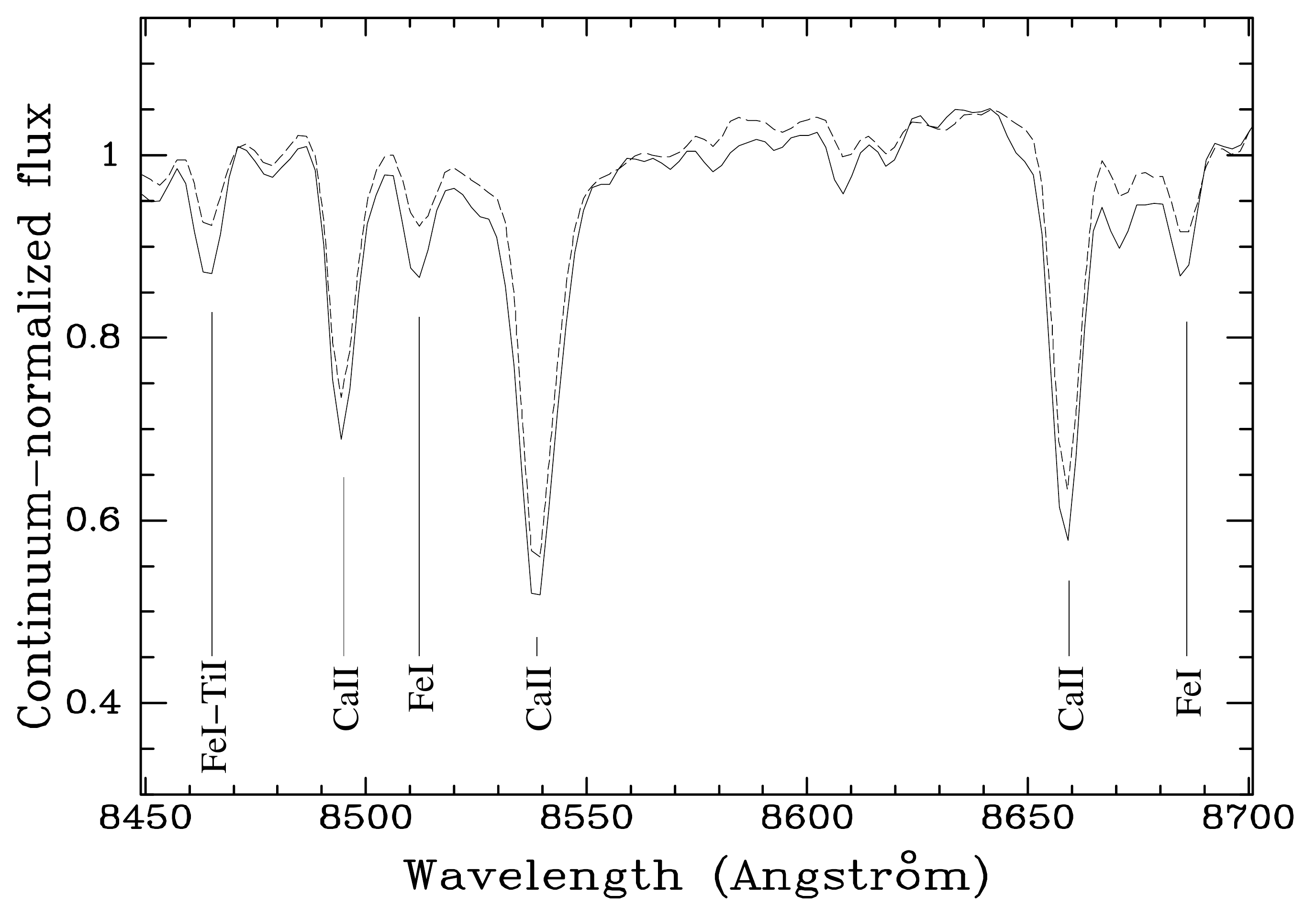}
\caption []{Luminosity effects at spectral type M2 in the region around the calcium infrared triplet. The solid line corresponds to the M2Ib standard HD~10465, and the dashed line to the M2III standard HD~219734 \citep{Garcia89}. Both spectra have been ratioed to their best-fitting 7th-degree polynomial (fit to the whole $6000 < \lambda (\AA ) < 9000$ range) to facilitate comparison. The spectral lines used in this work as luminosity class discriminants are indicated.}\label{lum_effect_comp}
\end{center}
\end{figure}

\begin{figure}[ht]
\begin{center}
\hspace{-0.5cm}
\includegraphics [width=9cm, angle={0}]{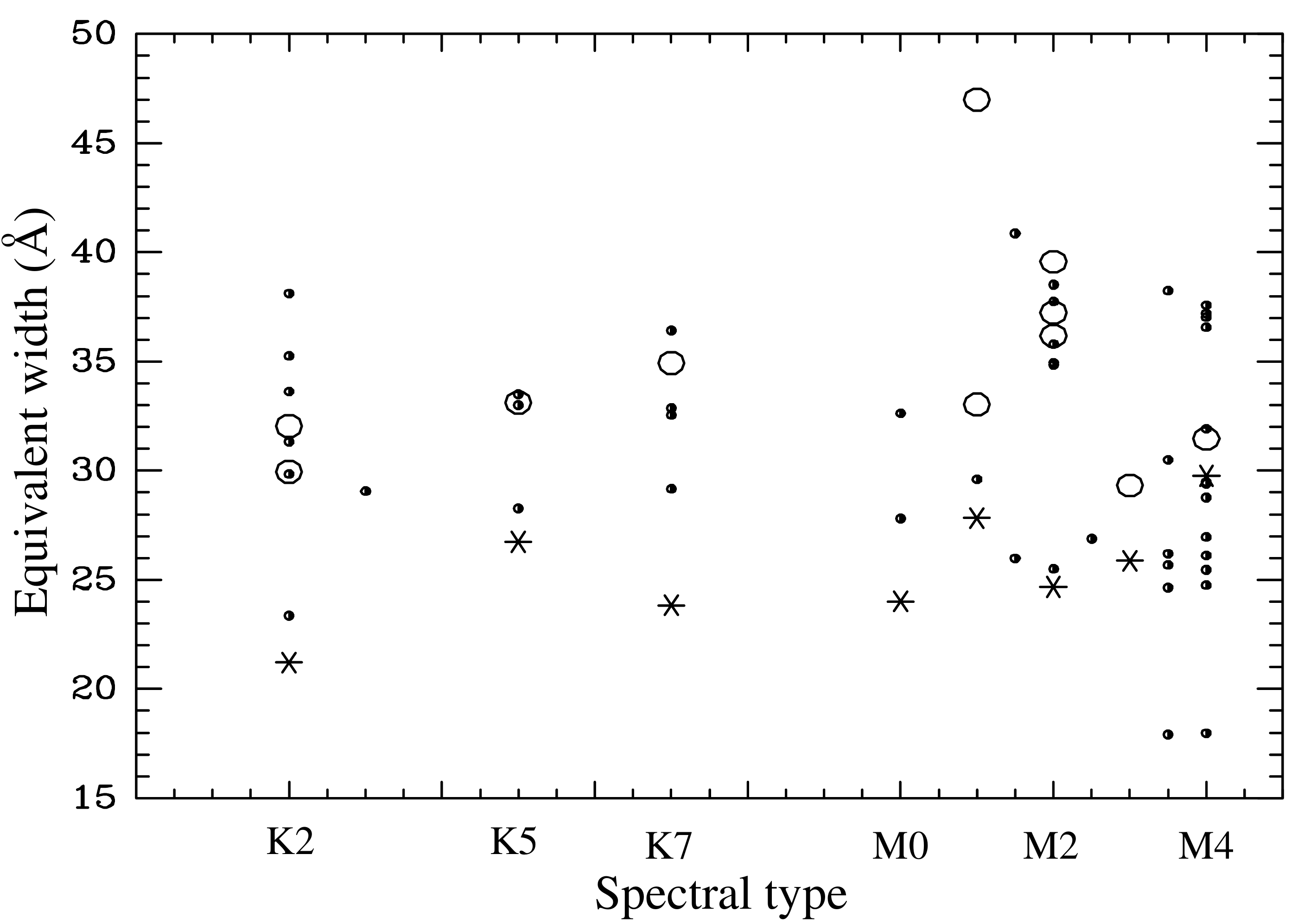}
\caption []{Added equivalent widths of all the features used as luminosity class discriminants as indicated in Sect.~\ref{observations} as a function of spectral type. Open circles indicate the stars from \citet{Garcia89} used as standards of luminosity classes I and II, and asterisks correspond to standards of luminosity class III. The location of all the stars in Cygnus observed by us are represented by dots.}\label{indices_all}
\end{center}
\end{figure}

The distinction between giants and supergiants in the observed spectral range is based on rather subtle differences in spectral features displaying a dependency on surface gravity. Useful features for spectral classification in the red/far-red range across the entire spectral sequence are presented in detail in the atlas of \citet{Torres93}, which forms the basis of our luminosity classification. The most prominent luminosity-sensitive features in the K-M range are the cluster of CN bands near 7970~\AA\ and the CaII infrared triplet at 8498, 8542, and 8662~\AA , all of which have a positive luminosity effect with their equivalent widths increasing with luminosity. Other features that we used are the blend of metallic lines at 6497~\AA , the atomic lines of TiI in the 7345-7364~\AA\ interval, and the KI line at 7699~\AA\ , also having a positive luminosity class effect. To these we added other luminosity-sensitive features discussed by \citet{Keenan45} due to FeI-TiI at 8498~\AA\, and FeI at 8514 and 8689~\AA\ , again having a positive luminosity effect (see Fig.~\ref{lum_effect_comp}). We measured equivalent widths of each of these features, both for the MK standards and for our target stars. For the standards, the equivalent width of any given feature as a function of spectral type traces a well-defined curve generally increasing from early to late K-types, and then reaching a plateau at late K- or early M-types, as shown in Fig.~\ref{indices_all}. The behavior is less systematic among the supergiants of luminosity classes Ia to II, for which the equivalent widths do not follow such a strict relationship. However, they always fall above those of giants of the same spectral type. Therefore, even though our spectra do not allow us to discern among the supergiants of luminosity classes Ia, Iab, Ib or II, the sequence of equivalent widths as a function of spectral type for luminosity class III giants forms a clear lower boundary, with supergiants generally having equivalent widths distinctly above that boundary.

We used the luminosity-sensitive features noted above to independently classify each star as giant or supergiant on the basis of each of them, and then we compared the results obtained across those various luminosity indicators. Most of the stars in the sample are consistently classified as giants or supergiants according to each of the luminosity-sensitive features used for classification. For some stars we found one or two of them for which the classification differs from the one obtained from the others, or for which the measured equivalent width is not sufficiently separated from the giant sequence locus to clearly classify them as giants or supergiants. We assigned those stars to the luminosity class obtained from the majority of the other lines. In a few cases in which the number of features favoring either classification was similar, we adopted the luminosity class derived from the CN and CaII features. Table~\ref{supergiants} lists our spectral classifications for the 29 supergiants identified in our sample.

\begin{figure}[ht]
\begin{center}
\hspace{-0.5cm}
\includegraphics [width=7cm, angle={0}]{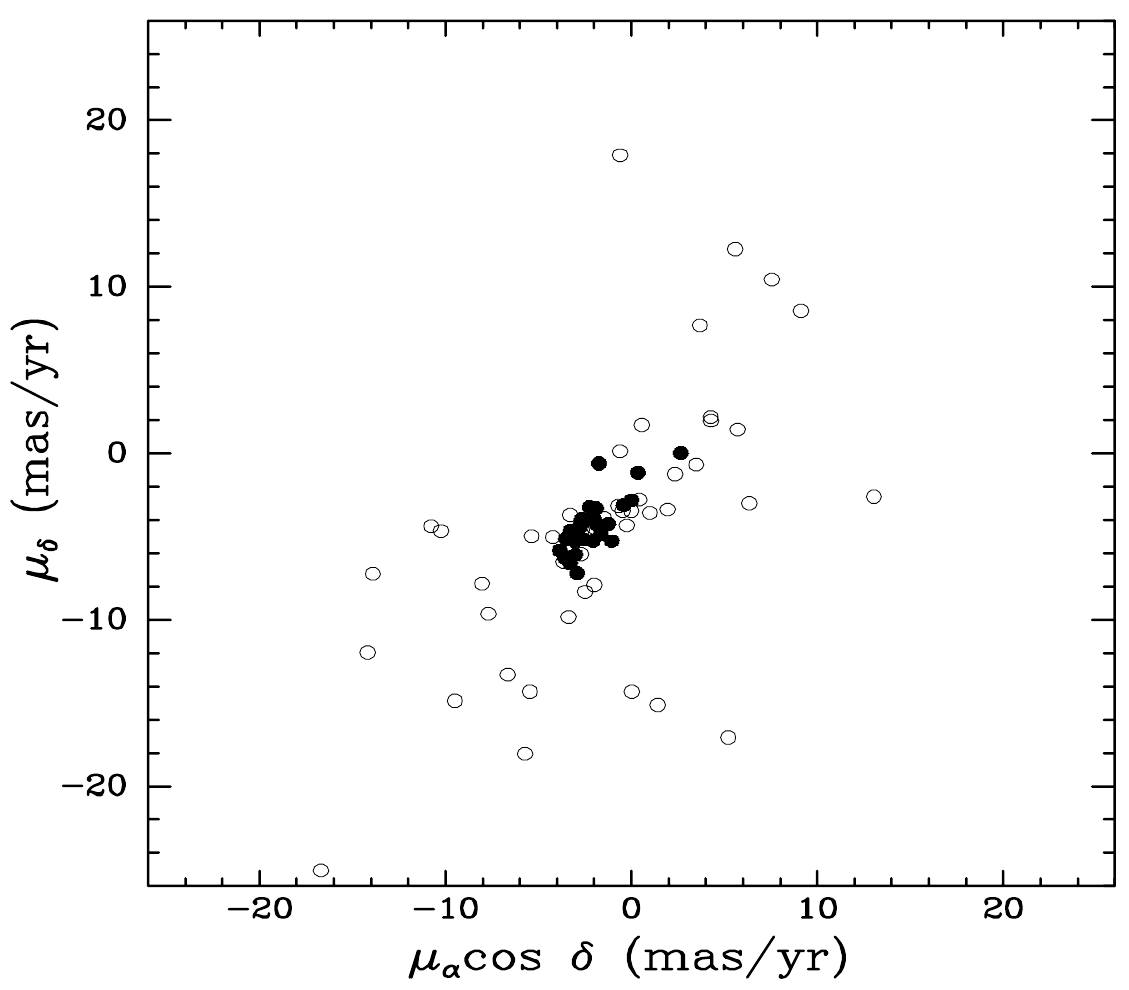}
\caption []{Gaia DR2 proper motions of the targets selected according to the criteria described in Section~\ref{target}. The 29 stars spectroscopically classified as supergiants are represented with filled circles, and the rest of the sample is represented by the open circles.}\label{proper_motion}
\end{center}
\end{figure}

\begin{figure}[ht]
\begin{center}
\hspace{-0.5cm}
\includegraphics [width=7cm, angle={0}]{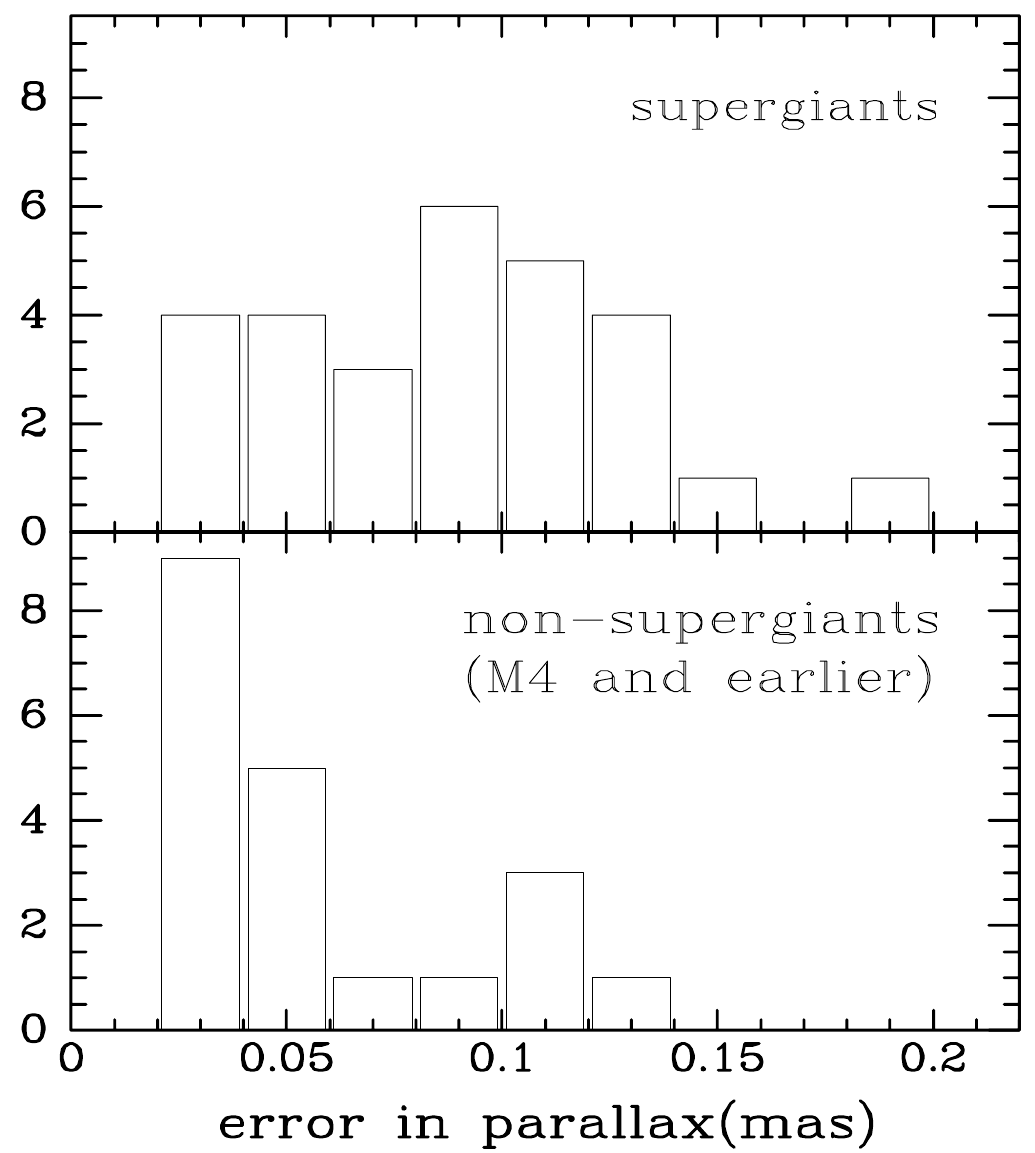}
\caption []{Distribution of the parallax errors listed in the Gaia DR2 catalog for the targets spectroscopically classified as supergiants in this work (upper panel) and those classified as giants covering the same range of spectral types. The effect of the intrinsic photospheric properties of supergiants on the accuracy with which parallaxes can be determined is apparent.}\label{errors}
\end{center}
\end{figure}

Figure~\ref{proper_motion} shows the distinct kinematical properties of the red supergiants, which display a low velocity dispersion, as compared to the proper motion distribution of all the other stars in our sample, which is characteristic of an evolved, kinematically hot population. The renormalized unit weight error (RUWE) quantifying the goodness of the astrometric solution \citep{Lindegren18} is well below the threshold of 1.4 for all the red supergiants, implying that their astrometric solutions provide good fits to the data, and only one, the M4 supergiant J202308.60+365145.0 with a RUWE of 1.366, is near that value. Somewhat surprisingly even J203323.91+403644.2, for which a negative parallax is obtained, has a RUWE of 0.997 indicative of a good astrometric solution. However, the precision with which parallaxes can be determined shows a clear distinction between supergiants and non-supergiants. Figure~\ref{errors} shows that astrometric solutions for red supergiants yield larger uncertainties in their parallaxes than for red giants in the same range of spectral types. This is expected from the higher noise of their astrometric solutions, as noted in Section~\ref{distance}.

Most of the stars in our sample are identified as supergiants for the first time. Twelve of the 29 supergiants had no previous spectral classification, and five had spectral types published in the literature but no luminosity class determined. Eleven have been classified as supergiants in previous works (see Notes to Table~\ref{supergiants}), with spectral types very close to the ones that we determine. As noted in Table~\ref{supergiants} one of the stars that we classify as M3.5 supergiant, J204924.95+443349.1, was included in previous studies as a candidate carbon star, which is not supported by our classification.

The membership of five of the 29 supergiants in the area of our study is unlikely, and we list them separately in Table~\ref{supergiants}. Two of them, J204636.65+424421.6 and J204924.95+443349.1, have large negative radial velocities, small parallax values, and their $K_S$ magnitudes are among the faintest in the sample, all of which is more consistent with membership in the background Perseus arm. This is also the case for J201126.21+354227.7, for which no radial velocity measurement is provided in the Gaia~DR2 catalog, but its measured parallax is the lowest among those in our sample, with the exception of J203323.91+403644.2 for which Gaia~DR2 lists a negative parallax. On the other extreme, we consider J203817.18+420425.1 as a likely K2 foreground supergiant, given its large parallax that places at a roughly half of the distance adopted for the Cygnus region. Finally, J202412.43+411402.9 is a M4 supergiant also with a large negative radial velocity consistent with membership to the Perseus arm, although unlike in the two cases previously discussed this is not additionally supported by a small parallax.

Although the vast majority of the remaining 24 supergiants are consistent with membership in the Cygnus region, a few doubtful cases subsist. J201919.97+341101.0 and J202736.78+385046.1 have proper motions deviant from the bulk of the sample and they may be runaway stars, although their radial velocities are within the range defined by the rest of the group. The difference in spatial velocity between J201919.97+341101.0 and the bulk of the stars in the region is approximately 50~km~s$^{-1}$, which is on the low end of typical runaway velocities, and the difference is even lower for J202736.78+385046.1. Finally, we have included among the Cygnus region members three M supergiants with cataloged parallaxes above 1~mas, J203301.05+404540.4, J203401.43+422526.6 and J203936.52+391209.7. We note that the uncertainty in their parallax measurements are among the highest in our sample, making them particularly unreliable.

\citet{Garmany92} listed BC~Cyg, BI~Cyg, KY~Cyg, and IRAS~20193+3527 (=J202114.07+353716.5) as possible M supergiant members of Cygnus associations, all of which we confirm. Two other M supergiant candidate members, IRAS~20212+3920 and IRAS~20002+3530, were not included in our sample because they were fainter than our magnitude limits, and they probably are background. A seventh candidate, RW~Cyg, was also excluded from our sample as likely foreground because of its brightness $(K_S = 0.48$ in 2MASS). From the candidate supergiants near Cygnus~OB2 in \citet{Comeron16} we confirm J203323.91+403644.2 and J203301.05+404540.4, although with somewhat earlier spectral types than those obtained in that work from $K-$band spectroscopy. We do not confirm as supergiants RAFGL~2600 (J203128.70+403843.3), for which we find a spectral type much later than M4, and IRAS~20249+4046 (=J202643.03+405626.8), which was already flagged as doubtful in \citet{Comeron16} due to their moderately deep CO bandheads.

\begin{table}
\caption{\label{giants}Late-type giants}
\begin{tabular}{lc | lc}
\hline\hline
\noalign{\smallskip}
Star & Type & Star & Type \\
\noalign{\smallskip}\hline\noalign{\smallskip}
J195955.80+351747.2 & M1   & J202710.20+350549.9 & M6.5 \\
J200138.41+360629.9 & M5.5 & J202733.62+440306.9 & M6.5 \\
J200339.49+381938.3 & M2   & J202817.90+371603.1 & M7   \\
J200513.51+390008.9 & M5.5 & J202828.53+374721.8 & M4   \\
J200707.30+360030.8 & M5.5 & J202848.11+372329.1 & M6   \\
J200846.88+360852.1 & M1.5 & J202850.59+395854.3 & M5   \\
J200932.98+334053.9 & M0   & J202856.38+353226.3 & M2.5 \\
J201213.90+355847.9 & M5   & J203128.96+423109.6 & M5.5 \\
J201228.80+332222.1 & M7   & J203234.06+443405.6 & M4   \\
J201240.08+390133.7 & M5   & J203425.45+452354.0 & M4   \\
J201313.38+364030.3 & M6   & J203427.55+465004.0 & M6   \\
J201313.63+404904.4 & M5   & J203436.00+410554.1 & K2   \\
J201337.51+350502.5 & M6.5 & J203518.01+423238.0 & M7   \\
J201613.84+383654.9 & M7   & J203555.61+373726.7 & M3.5 \\
J201726.47+343109.9 & M7   & J203758.15+423743.1 & M6   \\
J201814.21+323706.6 & M3.5 & J203813.03+394125.9 & M3.5 \\
J201830.76+350156.4 & K5   & J204111.07+460846.1 & M3.5 \\
J201936.82+334734.9 & M6   & J204715.58+405918.8 & M6.5 \\
J202132.42+392756.4 & M5   & J204845.10+405923.9 & M7.5 \\
J202243.84+434303.5 & M4   & J205352.82+442401.5 & M6.5 \\
J202448.01+425725.7 & M4   & J205418.34+423834.7 & M4   \\
J202508.81+441008.8 & M6   & J205552.09+422432.0 & M6   \\
J202518.27+434055.9 & M6.5 & J205741.53+433555.5 & M6   \\
\hline
\end{tabular}
\end{table}

As expected, almost all the other stars that met the selection criteria described in Section~\ref{target} turned out to be K and M giants, many of them with late spectral types. We classified them according to the MK spectral sequence in \citet{Fluks94}, and the results are presented in Table~\ref{giants}. Another star, J200743.20+392138.2, was found to be a reddened F0I supergiant. Finally, the only late-type star in our sample having a spectral type different from K or M is J205443.1+421825.5, a star previously identified as a long-period variable \citep{Wozniak04} with H$\alpha$ emission \citep{Kohoutek99}. Its spectrum is dominated by bands of ZrO, clearly identifying J205443.1+421825.5 as an S star. Our observations do not cover the telltale technetium lines at 4262\AA\ and 4238\AA\ that are one of the discriminating criteria between intrinsic and extrinsic S-type stars \citep{Jorissen92,Groenewegen93}, but its classification as a long-period variable and high luminosity derived from its Gaia~DR2 parallax ($\tilde \omega = 0.3922 \pm 0.1201$) strongly suggest that J205443.1+421825.5 belongs to the intrinsic type. S-type stars are evolved stars in either scenario, and we will not consider it further in the context of this work.

\section{Results\label{results}}

\subsection{Intrinsic properties\label{physpars}}

We adopt the spectral type versus effective temperature ($T_{\rm eff}$) calibration for galactic supergiants from \citet{Levesque07} to obtain $T_{\rm eff}$ for the stars in our sample. We estimate then the $K_S$-band extinction using the $T_{\rm eff}$ versus intrinsic $(J-K)_0$ color relationship for supergiants from \citet{Kucinskas05}, and the color transformation from the Johnson-Glass photometric system used by \citet{Kucinskas05} to the 2MASS photometric system using the transformation equations from \citet{Carpenter01}:

\begin{equation}
(J-K_S)_{\rm 2MASS} = 0.983 (J-K)_{\rm JC} - 0.018.
\end{equation}

The $K_S$-band foreground extinction $A_K$ is obtained using the extinction law of \citet{Cardelli89} with a total-to-selective extinction ratio $R_V = 3.1$ as

\begin{equation}
A_K = 0.686 [(J-K_S) - (J-K_S)_0].
\end{equation}

Finally the luminosity $L$ is obtained using the $T_{\rm eff}$ vs. bolometric correction $BC_K$ for galactic supergiants from \citet{Levesque05} and the adopted distance modulus $DM = 10.9$ (see Sect.~\ref{distance}):

\begin{equation}
\log (L / L_\odot) = -0.4 (K_S - DM - A_K + BC_K - 4.74).
\end{equation}

Once $T_{\rm eff}$ and $L$ are derived as described, we estimate the initial mass and the age of each star using the Geneva evolutionary models \citep{Ekstroem12}. We base our estimates on the rotating models with solar metallicity, and an initial rotation velocity that is 40\% of the critical value, which we take as being representative of the ensemble. In the mass interval relevant for our sample, rotation at that initial velocity increases $\log L$ by $\sim 0.1$~dex in the red supergiant phase with reference to the nonrotating models, which is of the same order as the variation of the luminosity during that phase. Likewise, rotation increases the age at which the helium burning marking the beginning of the red supergiant phase is reached by $\sim 20$\% with respect to the nonrotating case. We adopt the duration of the red supergiant phase as the duration of the helium-burning stage or, for stars with mass below $\sim 12$~M$_\odot$ undergoing a blue loop during that stage, the time span between the beginning of the helium burning and the beginning of the blue loop, since the post-blue loop helium-burning stage is much shorter than the pre-blue loop one. The duration of the red supergiant phase defined in this way in the rotating case increases from 1.8~Myr for an initial mass of 7~M$_\odot$ to 2.2~Myr for an initial mass of 10~M$_\odot$, then decreasing to 0.3~Myr at 20~M$_\odot$.

\begin{table*}
\caption{\label{physpar}Derived intrinsic properties of supergiants in the Cygnus associations}
\centering
\begin{tabular}{lcccccccc}
\hline\hline
\noalign{\smallskip}
Star & type & $T_{\rm eff}$ & $(J-K_S)_0$ & $BC_K$ & $A_K$ & $\log {L \over L_\odot}$ & M           & $t_{\rm RSG}$ \\
     &      & (K)           &             &        &       &                          & (M$_\odot$) & (Myr)               \\
\noalign{\smallskip}\hline\noalign{\smallskip}
J201842.17+411023.7 & K3   & 3970 & 0.93 & 2.56 & 1.144 & 4.175 & 10 & 25 \\
J201848.17+423637.4 & K7   & 3815 & 0.97 & 2.67 & 0.485 & 4.089 & 10 & 28 \\
J201857.52+390015.1 & K5   & 3840 & 0.96 & 2.65 & 0.224 & 4.517 & 12 & 19 \\
J201905.19+394715.9 & K7   & 3815 & 0.97 & 2.67 & 0.482 & 4.002 &  9 & 33 \\
J201917.19+382550.2 & K2   & 4015 & 0.88 & 2.52 & 0.227 & 3.763 &  8 & 47 \\
J201919.97+341101.0 & K2   & 4015 & 0.88 & 2.52 & 0.220 & 3.775 &  8 & 46 \\
J202114.07+353716.5 & M2   & 3660 & 1.05 & 2.81 & 0.418 &  4.65 & 13 & 17 \\
J202121.92+365555.5 & M4   & 3535 & 1.11 & 2.89 & 0.350 & 4.951 & 17 & 12 \\
J202133.14+363654.4 & K2   & 4015 & 0.88 & 2.52 & 0.798 & 4.015 &  9 & 32 \\
J202138.55+373158.9 & M4   & 3535 & 1.11 & 2.89 & 0.485 & 5.175 & 19 &  9 \\
J202211.42+405121.2 & K2   & 4015 & 0.88 & 2.52 & 0.595 & 4.155 & 10 & 25 \\
J202308.60+365145.0 & M4   & 3535 & 1.11 & 2.89 & 0.849 & 4.398 & 11 & 21 \\
J202502.94+433605.9 & M2   & 3660 & 1.05 & 2.81 & 0.481 & 3.784 &  8 & 46 \\
J202558.05+382107.6 & M4   & 3535 & 1.11 & 2.89 & 0.611 & 5.176 & 19 &  9 \\
J202603.31+410827.6 & M1.5 & 3710 & 1.03 & 2.75 & 0.769 & 4.498 & 12 & 19 \\
J202645.17+381343.5 & K5   & 3840 & 0.96 & 2.65 & 0.513 & 4.104 & 10 & 27 \\
J202736.78+385046.1 & K7   & 3815 & 0.97 & 2.67 & 0.431 & 4.127 & 10 & 25 \\
J203301.05+404540.4 & M2   & 3660 & 1.05 & 2.81 & 1.037 & 5.029 & 18 & 11 \\
J203323.91+403644.2 & K7   & 3815 & 0.97 & 2.67 & 1.363 & 4.135 & 10 & 25 \\
J203401.43+422526.6 & M3.5 & 3542 &  1.1 & 2.88 & 1.254 & 5.201 & 21 &  9 \\
J203936.52+391209.7 & M4   & 3535 & 1.11 & 2.89 & 1.179 & 4.825 & 15 & 14 \\
J204340.64+442837.6 & M2   & 3660 & 1.05 & 2.81 & 0.897 & 4.512 & 12 & 19 \\
J204604.55+445209.4 & K7   & 3815 & 0.97 & 2.67 & 0.367 & 4.356 & 11 & 21 \\
J205425.72+431247.0 & M0   & 3790 & 0.98 & 2.69 & 0.446 & 3.844 &  8 & 41 \\
\hline
\end{tabular}
\end{table*}

\begin{figure}[ht]
\begin{center}
\hspace{-0.5cm}
\includegraphics [width=8cm, angle={0}]{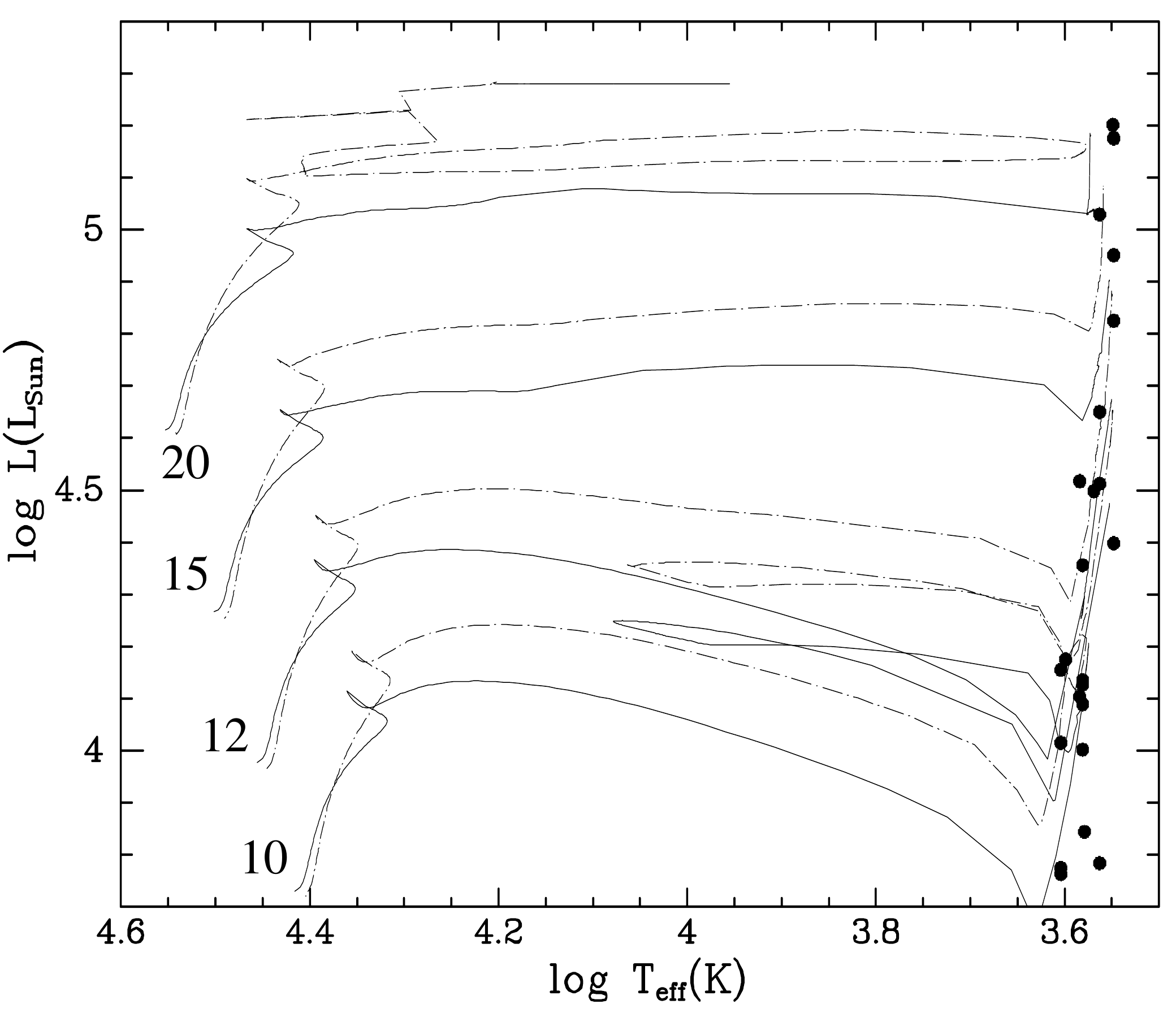}
\caption []{Derived temperatures and luminosities of the stars spectroscopically classified as supergiants and assigned to the Cygnus region, assuming a common distance module $DM=10.9$, plotted together with the evolutionary tracks of \citet{Ekstroem12} for stars with various initial masses. The solid lines describe the evolution of nonrotating stars, and the dot-dashed lines correspond to stars with an initial rotation of $40$\% of the critical value. Adopting a distance of 1.7~kpc, seemingly favored by Gaia, would shift the location of the stars upward by 0.1~dex in luminosity.}\label{teff_lum}
\end{center}
\end{figure}

Table~\ref{physpar} shows the intrinsic properties of the 24 red supergiants that we assign to the Cygnus region under those assumptions. The mass is derived from the average value of the luminosity during the helium burning phase in the rotating models, and $t_{\rm RSG}$ is the age at which the star enters the red supergiant phase. Figure~\ref{teff_lum} shows their location in the temperature-luminosity diagram, with the evolutionary tracks for rotating models of various initial masses. Most of the stars are located within the steep, narrow band defining the helium-burning sequence, with the latest spectral types generally near the upper part of that sequence. There are a few exceptions nevertheless, most notably J205425.72+431247.0 (M0) and especially J205425.72+431247.0 (M2). Both stars appear too faint for their spectral type, and their parallax, proper motion, radial velocity and derived extinction are well within the range covered by other red supergiants that we assign to the Cygnus region. Their classification as supergiants is firmly established, as they meet each one of the luminosity class criteria discussed in Section~\ref{observations}, and also the spectral subtype is well determined from the shape of the temperature-sensitive molecular features. The difference in derived luminosity with respect to other members of the Cygnus region having similar spectral types reaches nearly one order of magnitude, which would place them in the Perseus arm at a distance of $\sim 4.2$~kpc if the true luminosity were typical of their spectral type. We consider this possibility very unlikely, as it would imply that the parallax should be in error by more than $3 \sigma$ for both stars. Furthermore, their radial velocities differ by $\sim 40$~km~s$^{-1}$ from the typical radial velocities found in the Perseus arm members (like J204636.65+424421.6 and J204924.95+443349.1; see Table~\ref{supergiants}). No previous studies are found in the literature about these two stars that could help to elucidate the reason of their discrepant position in the $\log T_{\rm eff}$ versus $\log L$ diagram.

\subsection{Spatial distribution\label{spatial_dist}}

The age range covered by red supergiants makes them poor tracers of the location of their parent associations. Taking the velocity dispersion inside a typical OB association as 4.5~km~s$^{-1}$ \citep{Melnik20}, red supergiants can be expected to have moved typically by $\sim 100$~pc from their birthplaces in 20~Myr, a distance exceeding by a factor of a few the typical size of an OB association. Therefore, we should not expect to find our red supergiants confined to the boundaries of the identified OB associations.

\begin{figure*}[ht]
\begin{center}
\hspace{-0.5cm}
\includegraphics [width=14cm, angle={0}]{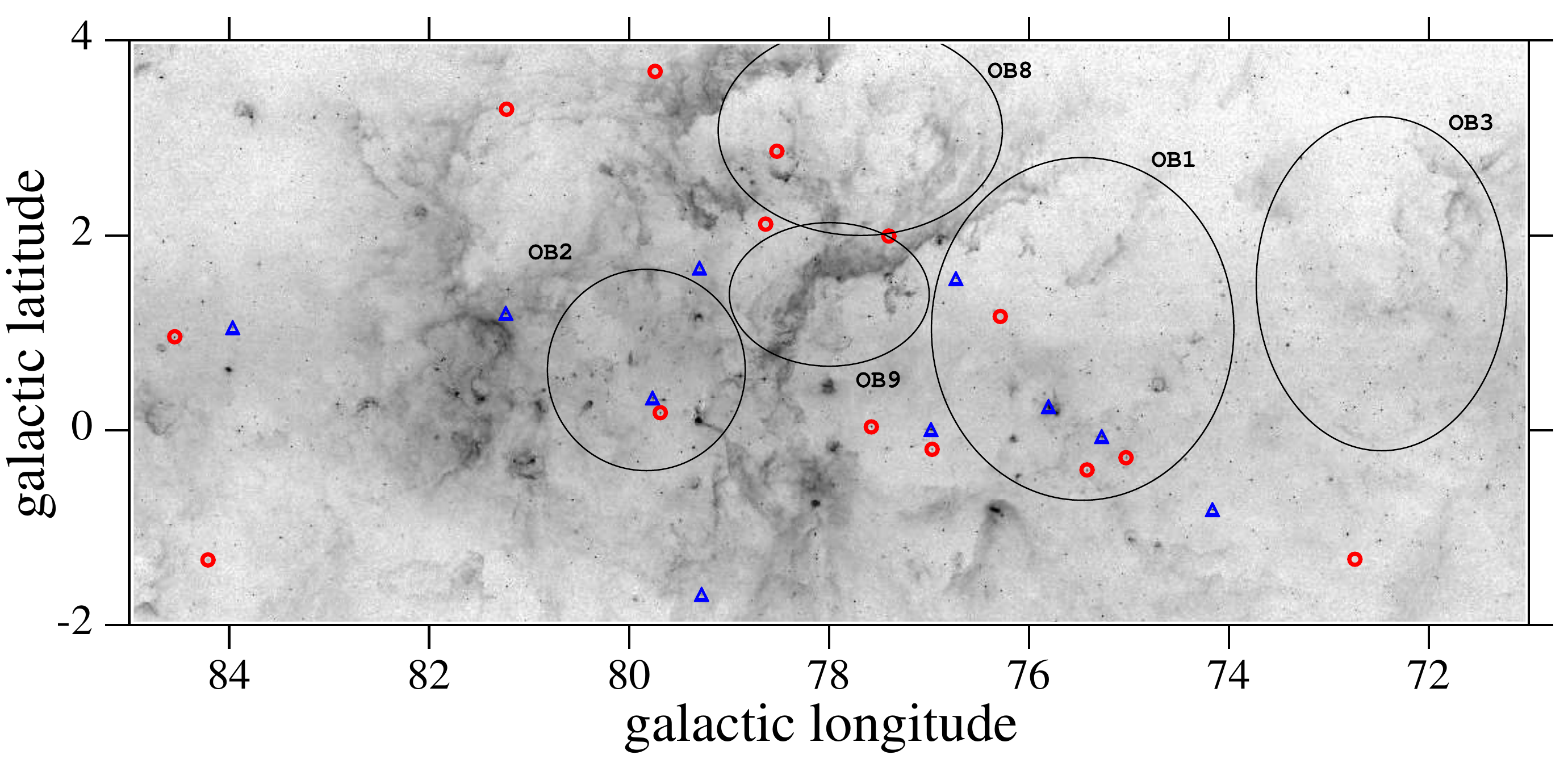}
\caption []{Spatial distribution of the red supergiants belonging to the Cygnus region, overplotted on a mosaic of infrared images obtained by the Midcourse Space Experiment (MSX) satellite in the infrared band $A$ ($6.0-10.9~\mu$m) that outlines the distribution of warm dust. Blue triangles indicate the positions of stars younger than 20~Myr, and red circles those of older stars. The approximate boundaries of present-day associations as listed by \citet{Uyaniker01} are indicated. The pervading nebulosity between $l \simeq 77^\circ$ and $l \simeq 83^\circ$ outlines the Cygnus~X molecular complex.}\label{distr}
\end{center}
\end{figure*}

Figure~\ref{distr} shows the position of the red supergiants in the region under discussion, superimposed on a map obtained by MSX, the Midcourse Space Experiment satellite \citep{Price95} in the infrared band $A$ ($6.0-10.9~\mu$m) that highlights the interstellar medium as traced by warm dust. Red supergiants show a very mild concentration toward the central regions of the map, broadly overlapping the area covered by the associations Cygnus OB1, OB2, OB8, and OB9 although some stars appear outside those boundaries. The three stars with the highest galactic longitude (near the left border of Figure~\ref{distr}) are within the boundaries of Cygnus~OB7, which extends further toward higher galactic longitude and is at roughly half of the distance to the other OB associations. However, there is no clear evidence supporting membership of these stars to Cygnus~OB7, as both their Gaia~DR2 parallaxes, proper motions and radial velocities are within the range of the other stars that we consider members of the Cygnus region.

Figure~\ref{distr} also shows the relative positions of red supergiants with estimated ages above and below 20~Myr. No obvious trend is found, ruling out the existence of a large-scale age gradient traced by them. Indications of a progression in star formation with time from lower to higher latitudes has been reported in the surroundings of Cygnus~OB2 \citep{Drew08,Comeron08,Comeron12} based on younger stars. The lack of evidence for a large-scale gradient among the red supergiant population suggests that the progression reported is probably a localized and more recent feature, where massive star formation initiated in Cygnus~OB9 propagated toward Cygnus~OB2, and currently continues at even higher longitude in Cygnus~X North, where the DR21 complex is an active site of ongoing massive star formation \citep{Motte07,Comeron08,Csengeri11}.

There is an absence of red supergiants at the lowest galactic longitudes, with no such stars appearing within or anywhere near the boundaries of the relatively rich association Cygnus~OB3 \citep{Rao20}, with the only exception of the M4 supergiant 201126.21+354227.7 (V430~Cyg). This star is listed in Table~\ref{supergiants} as having a parallax $\tilde\omega = 0.2385 \pm 0.0779$~mas, which is the lowest nonnegative parallax of our sample, therefore being most likely a background object. Recent Gaia~DR2-based results on Cygnus~OB3 itself suggest a distance of $2.0 \pm 0.3$~kpc \citep{Rao20}, somewhat more distant than the Cygnus~OB1/OB2/OB9 complex \citep{Straizys19,Rao20}. However, our criteria used to select red supergiant candidates should have selected them as well, as the modest increase of 0.6~mag in distance modulus should be partly offset by a lighter extinction \citep{Straizys19} in its direction. The lack of young red supergiants might be due to massive star formation in the region having started only recently, implying that only the most massive stars, with very short lifetimes as red supergiants, have entered that phase. The catalog of \citet{Kharchenko16} lists seven possible open clusters in the area, five of them with estimated ages above 10~Myr, and their distances are consistent with those of the Cygnus region \citep{Cantat18}. However, the membership to most of them, and even their physical existence as clusters, is highly uncertain due to the crowdedness of the area. The most outstanding cluster in Cygnus~OB3, NGC~6871, is listed by \citet{Kharchenko16} as having an age just short of 10~Myr, still consistent with the youth of the area.

\subsection{Kinematics\label{kinematics}}

At the adopted distance of 1.51~kpc, the typical proper motion uncertainty of $0.2$~mas~yr$^{-1}$ in the Gaia DR2 catalog translates into a tangential velocity uncertainty of only $1.4$~km~s$^{-1}$, well below the internal velocity dispersion of individual associations \citep{Melnik20}. In turn, the standard deviation of the proper motions of the stars that we allocate to the Cygnus region translates into a standard deviation of the tangential velocity of 9.4~km~s$^{-1}$, which is of the same order as the standard deviation of the radial velocities, 7.1~km~s$^{-1}$. The internal velocity structure of OB associations and the relative velocities among groups is therefore well resolved by Gaia, and the fact that the dispersion in space velocity among the stars in our sample is significantly larger than the internal velocity dispersion in associations suggests that they had their origin in kinematically distinct groups across the region.

The kinematic accuracy, combined with a galactic potential model, allows us to investigate the regions of origin of our sample of red supergiants. Age is nevertheless the dominant source of uncertainty, followed by the distance. The spread of distances along the line of sight is likely to be relevant as well since, even if the stars might have been born within a narrow range of distances to the Sun, the velocity dispersion along the line of sight introduces a scatter reaching $\pm 300$~pc in distance for the ages of the oldest stars in our sample, of the same order as the uncertainty in the distance derived from the trigonometric parallax. Taken together these factors severely limit our ability to determine the location of origin of the red supergiants, especially of the oldest among them. However, some intriguing trends do appear.

We traced back in time the trajectories of the 23 red supergiants of the Cygnus region for which full kinematical information is available using an axisymmetric galactic potential that yields a flat rotation curve within the range of galacticentric distances covered by the orbits of the stars in our sample. The radial force $f_R$ toward the galactic center per unit mass is then

$$f_R = 1.0226 \times 10^{-9} {\rm km \ s^{-1} \ yr^{-1}} V_0^2 [{\rm km^2 \ s^{-2}}] R^{-1}[{\rm kpc^{-1}}], \label{fr}$$

\noindent where $V_0$ is the circular velocity of the local standard of rest velocity around the galactic center and $R$ is the distance to the galactic center. We adopt the parameters obtained by \citet{McMillan17}: $V_0 = 232.8$~km~s$^{-1}$, distance of the Sun to the galactic center $R_0 = 8.20$~kpc, and components of the velocity of the Sun with respect to the local standard of rest $(U_\odot, V_\odot, W_\odot) = (11.10, 12.24, 7.25)$~km~s$^{-1}$, where the three components of the velocity are directed toward the galactic center, the direction of galactic rotation, and the north galactic pole, respectively. We assume the force per unit mass perpendicular to the galactic plane $f_z$ to be proportional to the distance $z$ to it, and use the parameters that reproduce the force at 1.1~kpc from \citet{Kuijken91}, as also adopted by \citet{McMillan17}, obtaining

$$f_z = 9.284 \times 10^{-7} {\rm km \ s^{-1} \ yr^{-1}} z[{\rm kpc}]. \label{fz}$$

\begin{figure}[ht]
\begin{center}
\hspace{-0.5cm}
\includegraphics [width=9cm, angle={0}]{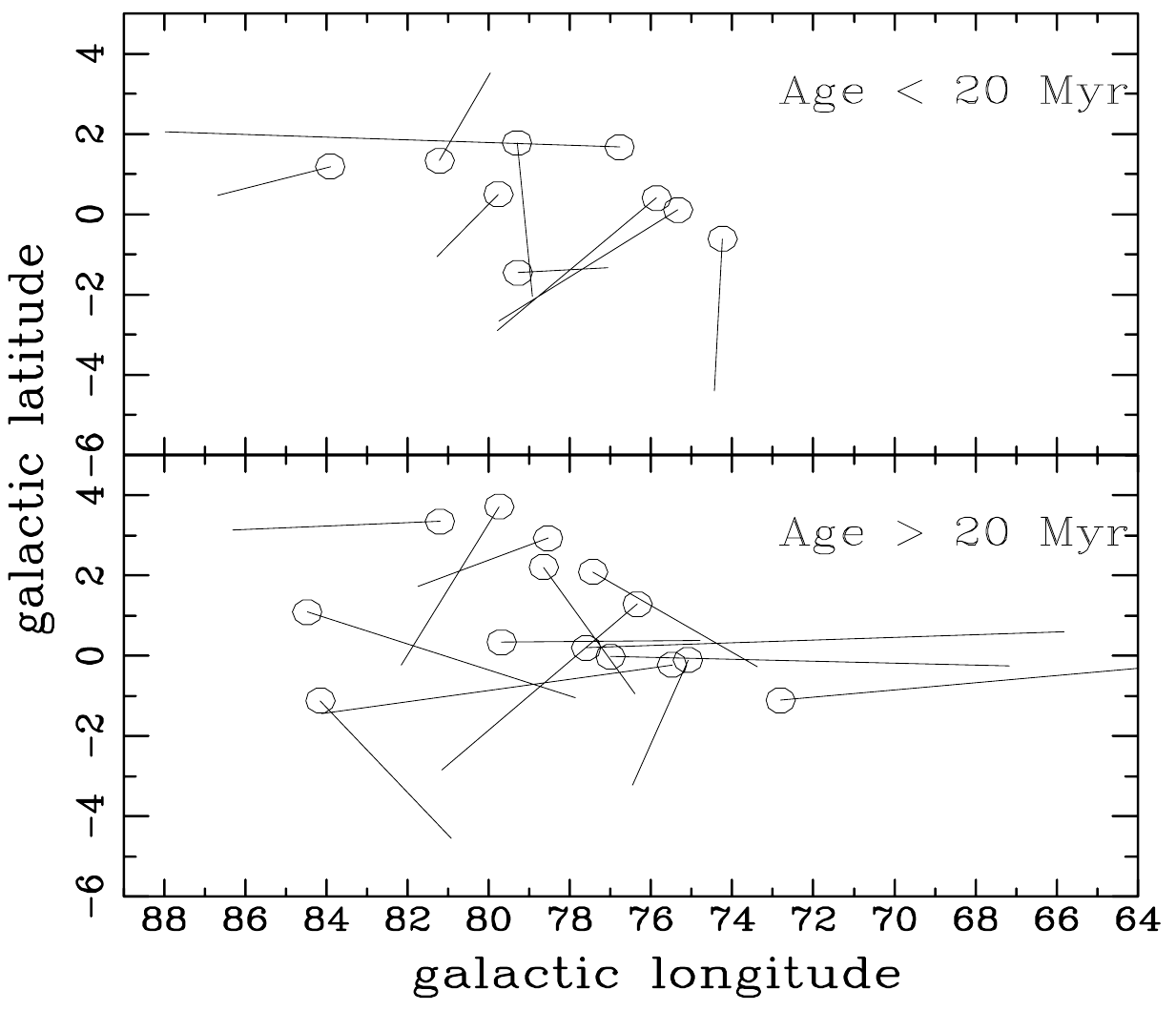}
\caption []{Positions at present (circles) and 15~Myr ago (end of the lines having their origin at the present position of each star) for the 23 red supergiants in Cygnus having measured proper motions and radial velocity. The sample is divided as in Figure~\ref{distr}, with the younger and older samples separated in two panels for clarity. The galactic longitude and latitude are defined as they would be measured by an observer moving with the circular velocity along the solar circle and located at the present time at the position of the Sun.}\label{thenandnow}
\end{center}
\end{figure}

We integrated the orbits of the 23 stars backward in time, also computing the evolution of their galactic coordinates as would be measured by an observer moving with the circular velocity along the solar circle at a position coinciding with that of the Sun at the present time. Figure~\ref{thenandnow} shows the present position of each star projected on the sky as compared with its position 15~Myr ago, which roughly gives the locations of the young stars (defined as those having ages estimated at less than 20~Myr from the evolutionary models with rotation) near the time when they formed.

As it can be seen in Figure~\ref{thenandnow}, although old and young stars occupy similar distribution on the sky at present, both groups are kinematically different on the average. Most of the stars in the young group would have appeared at somewhat higher galactic longitude for an observer 15~Myr ago, with only 3 out of the 9 stars in that group having been located at somewhat lower galactic longitude, differing by $2^\circ 1$ or less from their present position. This is in contrast with the location of the stars in the old group at the same time in the past: Out of 14 stars in the group, ten would have been observed at lower galactic longitude, and all ten would have a galactic longitude differing by more than $2^\circ$ from their current position. That is, the young and old groups, which at present overlap in the plane of the sky, would have been spatially separated 15~Myr ago, and the average galactic longitude of the young group would have been larger than that of most members of the old group.

\subsection{The star formation history in the Cygnus region\label{sfr}}

The distribution of initial masses of present-day red supergiants represents an historical record of the star formation history of the region, which we can reconstruct thanks to their relationship with their ages. We use the cumulative mass distribution of the 24 red supergiants that we identify in the Cygnus region, following a treatment similar to that presented by \citet{Comeron16} and \citet{Comeron18}. The number $N(M)$ of red supergiants observed at the present time $t = 0$ with a mass above a certain value $M$ can be written as

\begin{equation}
N(M) \propto \int_{t=-\infty}^{0} \int_{M'=M}^{\infty} SFR(t) V(M',-t) \Psi(M') {\rm d} M' {\rm d} t,
\label{Ncumul}
\end{equation}

\noindent where $SFR(t)$ is the star formation rate as a function of time, $\Psi(M')$ is the initial mass function that we assume to remain constant with time, and $V(M',-t)$ is a hat function that has unit value if a star of mass $M$ and age $-t$ is undergoing the red supergiant phase at present, and zero otherwise. We note that we define $t$ as negative toward the past, hence the negative sign in front of $t$ to denote the present age of the star. Also, it may be noted that $V(M,t)$ should also include a dependency on the initial rotation velocity of the massive precursor, and the expression for $N(M)$ should therefore include an additional integral over the initial rotation velocity and its distribution function. The latter has as general features a peak at small velocities, followed by a decrease and then a high rotation velocity tail  \citep{Huang10}. Since the initial rotation velocities of our stars are not known, we ignore such details in our treatment, as we did when computing the estimated physical properties of the stars in Table~\ref{physpar}, assuming instead a typical rotation velocity for all of them.

\begin{figure}[ht]
\begin{center}
\hspace{-0.5cm}
\includegraphics [width=8cm, angle={0}]{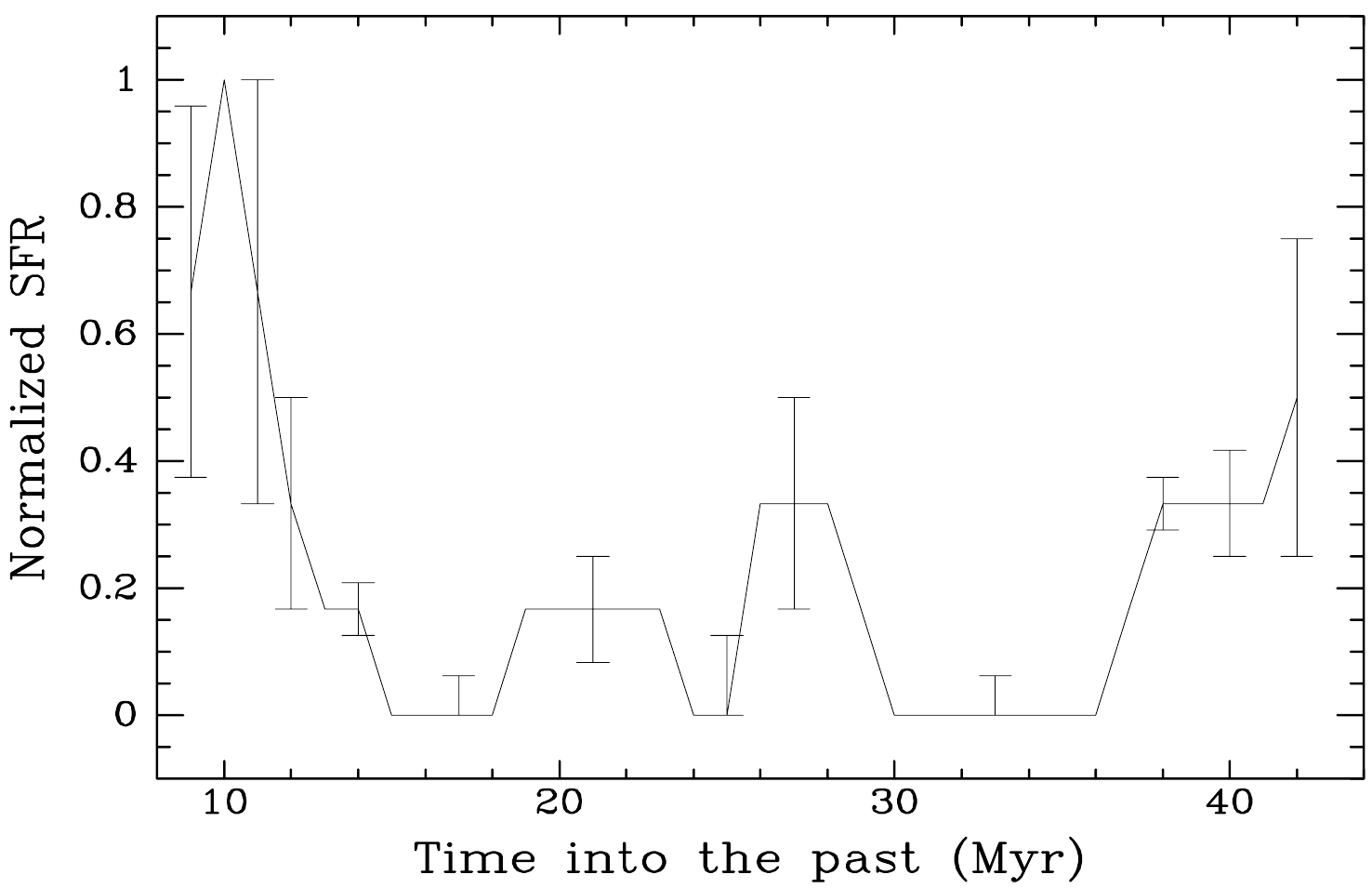}
\caption []{Reconstruction of the star formation rate normalized to its value 10~Myr ago, fit to reproduce the currently derived distribution of masses of red supergiants in Cygnus, assuming evolutionary models of \citet{Ekstroem12} with initial rotation $40$\% of the critical value. Error bars reflect uncertainties due to the actual initial rotation velocities and to the limited number of stars in each mass bin.}\label{sfr}
\end{center}
\end{figure}

Using a power law with Salpeter slope $-2.35$ in linear mass units for the initial mass function and $V$ from the evolutionary models of \citet{Ekstroem12}, we can numerically invert the integral. In this way we obtain the best fitting shape of the function $SFR(t)$ describing the overall star formation history traced by our sample.

\begin{figure}[ht]
\begin{center}
\hspace{-0.5cm}
\includegraphics [width=8cm, angle={0}]{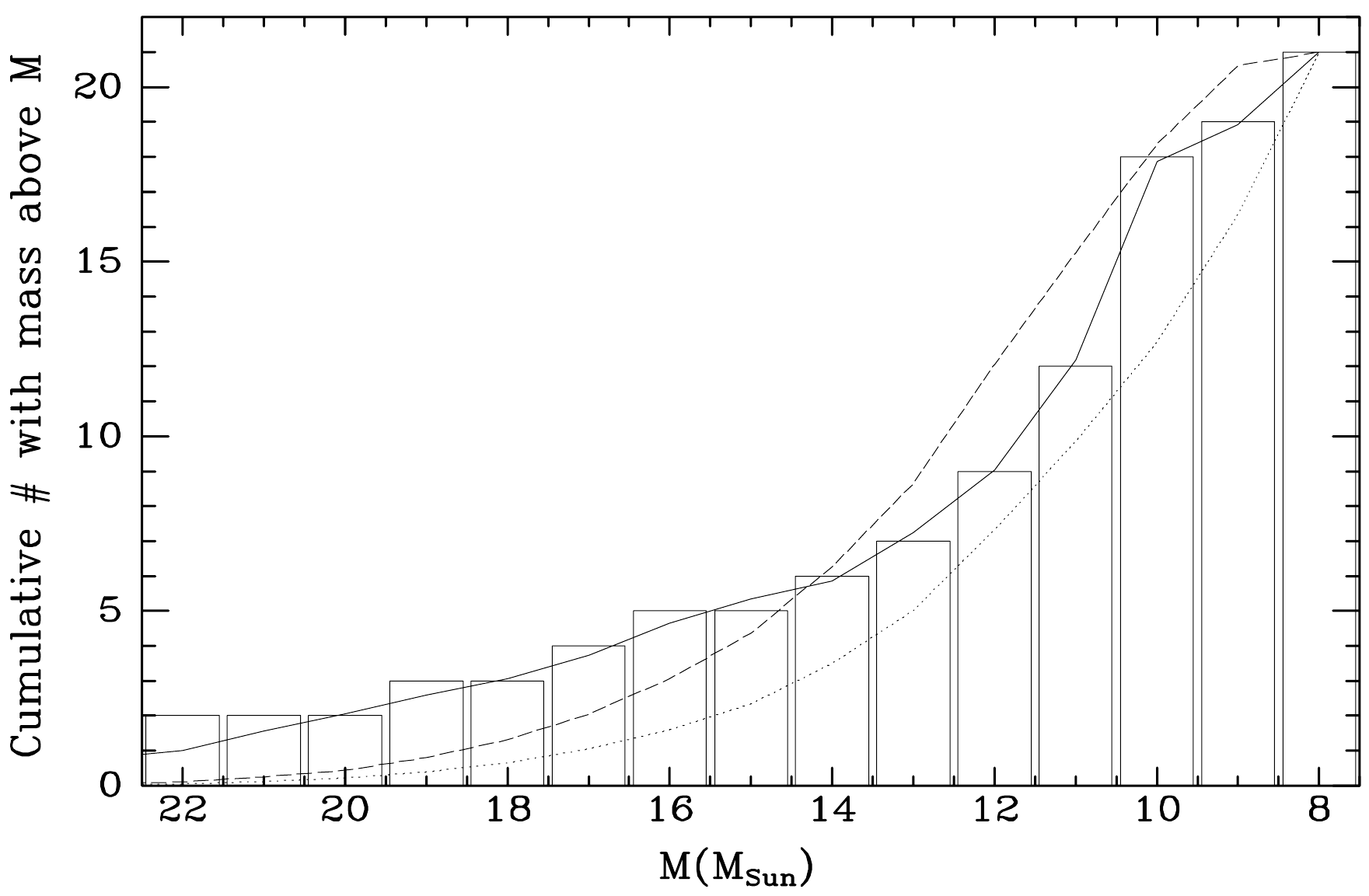}
\caption []{Cumulative histogram of masses of the 24 red supergiants in Cygnus identified in this work. The solid line shows the corresponding best-fit cumulative histogram derived from the star formation history depicted in Figure~\ref{sfr}. For reference, the predicted cumulative histograms obtained assuming a constant star formation rate (dotted line) and a star formation rate increasing linearly with time (dashed line) are shown as well.}\label{histo_cum}
\end{center}
\end{figure}

The star formation history $SFR(t)$ that we obtain in this way is shown in Figure~\ref{sfr}. A piecewise $SFR(t)$ function is adjusted by fitting its value at 1~Myr intervals so as to obtain the best possible match to the cumulative histogram of masses obtained from Table~\ref{physpar} through Eq.~\ref{Ncumul}. To estimate the uncertainty on $SFR(t)$ we have repeated the fit multiple times by adding Poissonian noise to the various mass bins to account for the uncertainties introduced by small number statistics, as well as by increasing or reducing at random the time of entrance of each star in the helium burning phase by up to 20\% to mimic the effects of the unknown initial rotation velocity of each individual star. The cumulative histogram is presented in Figure~\ref{histo_cum}. In this way, a star formation history in the region emerges in which the star formation rate started rising steeply only about 15~Myr ago. This is consistent with the estimated ages of the most prominent clusters of the region such as NGC~6871, NGC~6910, and NGC~6913, as well as with the beginning of the star formation in the present-day associations Cygnus OB1, OB8, and OB9. Such a recent increase in the star formation rate is required to account for the abundant presence of high-mass red supergiants and the rather shallow tail of the cumulative histogram of masses in the high-mass end. The star formation rate dropped to a very low level between approximately 15~Myr and 20~Myr ago, with no significant levels of high-mass star formation traced by the present-day red supergiants content.

Some sustained period of high-mass star formation took place over a span of $\sim 10$~Myr, starting about 30~Myr ago. O-type stars formed during this episode have completed their lifecycles and are not visible anymore, and the red supergiants that trace it come from B-type precursors currently observed as late K-type supergiants. Some open clusters with ages estimated to lie in this range are listed by \citet{Kharchenko16}, and are probably part of this episode. Despite being an active period of star formation, we estimate its intensity to have reached no more than $20-40$\% of the level of the peak period 10~Myr ago. This period was preceded by another period of inactivity, roughly 10 ~Myr long.

The lowest-mass red supergiants visible at present are the descendants of B1-B2 stars formed roughly 40~Myr ago. The dispersion in space of the stars formed at this time, the likely incompleteness of our sample at the lowest luminosities and the larger uncertainties in ages make it difficult to give any details on the star formation activity in that period, but the increase in the number of stars with initial masses near or below 10~M$_\odot$ requires it to have been rather intense, perhaps at the level of $\sim 50$\% of the peak level of 10~Myr ago, and more extended in time.

For the sake of comparison, Figure~\ref{histo_cum} also shows the histogram of masses with the histogram predicted by two simple star formation histories, in both cases normalized to the total number of stars. A constant star formation rate (dotted line in Figure~\ref{histo_cum}) consistently underpredicts the number of stars at all masses above the lowest-mass bin, or alternatively greatly overpredicts the number of low-mass red supergiants if the normalization condition is lifted. On the other hand, a star formation rate having started from zero 40~Myr ago and monotonically rising with a constant slope until 9~Myr ago, which is the estimated age of the youngest stars in our sample, yields a somewhat better match at higher masses (dashed line) but predicts a much higher number of the stars in the lowest mass bins, even at those where completeness of our magnitude-limited sample is not a concern. Clearly both simplified shapes of the star formation rate fail to reproduce the distribution of properties of the red supergiants in Cygnus.

\subsection{A possible link with the large-scale spiral structure\label{spiral}}

The fossil record of  massive star formation in Cygnus nowadays represented by red supergiants shows that the most intense activity started in the recent past (10-15~Myr ago), although older descendants from massive stars exist in the area at present. When the kinematics of the stars is considered jointly with the star formation history of the region we find that the latter cannot be interpreted simply in terms of a sequence of bursts in a giant molecular cloud complex. One of our most remarkable findings is the separation between most of the stars with ages below 20~Myr and most of the stars with ages above that value when their spatial locations are traced back 15~Myr or more in the past. This suggests that, although the giant molecular complex out of which the newest generations of stars have formed also harbors older stars, most of those stars formed elsewhere.

In a qualitative way, our results suggest that the giant molecular complex precursor of the present OB associations and active star forming sites was permeated over 10~Myr by a slow stream of massive stars coming from the general direction currently defined by its lower galactic longitude edge, and having formed over an extended period ranging from approximately 20 to more than 40~Myr. Those stars are now undergoing the relatively short-lived (less than 2~Myr) red supergiant phase, but they were early B-type stars when they encountered the Cygnus complex. At that point they were most likely accompanied by somewhat higher-mass stars that have already ended their lifecycles as supernovae in the meantime, whereas even higher-mass stars having formed together with them went through their entire evolution before encountering the Cygnus complex.

Since our results suggest that most of the stars currently observed as old red supergiants did not form within the present Cygnus molecular complex, it is interesting to speculate about their possible place of origin, perhaps linking it to the large-scale structure of our Galaxy and the history of the local environment in the last $\sim 50$~Myr. Thanks to the recent improvements in the determination of distances to tracers of spiral structure, especially through VLBI parallaxes to massive star forming regions \citep{Xu16,Reid19} and to young stars and aggregates with Gaia \citep{Chen19,Cantat20,Khoperskov20}, the existence of four major arms that can be followed from the inner regions of the Galaxy all the way to beyond the solar circle appears to be well established (see however \citet{Grosbol18} for a different view). In this picture the solar neighborhood is roughly halfway between two major arms, and the Cygnus region is the most prominent structure of the so-called Local Arm \citep{Xu13}, which may include the Orion complex as its nearest point to the Sun. The connection of the Local Arm with the grand design spiral structure and the underlying density wave is not clearly established though.

The objects that delineate the spiral structure are generally related to ongoing or very recent massive star formation, and therefore provide a snapshot of the present appearance of our Galaxy. Its evolution can nevertheless be investigated through the use of other tracers that can be followed further back in time, leaving the trail of the passage of the spiral structure through the galactic disk. A review of such studies has been presented by \citet{Vallee20}, who discusses the fundamental parameters of spiral structure of 24 nearby galaxies, including ours. In particular, the results of \citet{Vallee18,Vallee20} derive an angular velocity of the spiral pattern $\Omega_{\rm sp} = 19$~km~s$^{-1}$~kpc$^{-1}$ for the Milky Way, slower than the circular angular velocity in the solar circle.

With the distance of the Sun to the galactic center of $8.20$~kpc adopted in Section~\ref{kinematics}, an average galactic longitude $l=77^\circ$ and an average distance of $1.51$~kpc for the bulk of the population of old red supergiants in Cygnus, their galactocentric distance is $R = 8.0$~kpc, slightly inside the solar circle. For a flat galactic rotation curve with a rotation velocity $V_0 = 232.8$~km~s$^{-1}$, their average angular rotation velocity is $\Omega = 29.1$~km~s$^{-1}$~kpc$^{-1}$, implying a drift with respect to the overall spiral structure with the angular velocity $\Omega - \Omega_{\rm sp} = 10.1$~km~s$^{-1}$~kpc$^{-1}$. When comparing to the present location of the major spiral arms, this indicates that the old red supergiants are moving relative to them coming from the direction of the Sagittarius-Carina arm and going toward the Perseus arm.

At present, the Sagittarius-Carina arm intersects the galactocentric circle of radius $8.0$~kpc near the position of the Carina OB1 and OB2 associations, at an approximate distance of $2.0$~kpc from the Sun in the direction $l \simeq 285^\circ$ \citep{Melnik98,Reid19}. From this we obtain an angular distance $\lambda$ from the Sagittarius-Carina arm in the proximities of Carina OB1 to the Cygnus region measured from the galactic center $\lambda \simeq 24^\circ 5$, which translates into an arc of length $D = (2 \pi R \lambda) / 360^\circ \simeq 3.4$~kpc along the galactocentric radius of 8.0~kpc. At the drift angular speed of $10.1$~km~s$^{-1}$~kpc$^{-1}$ with respect to the spiral structure, the old red supergiants would have covered the length of that arc in a time $t = D/[(\Omega-\Omega_{\rm sp}) R] \simeq 43$~Myr, which is similar to the age of the oldest red supergiants of our sample.

We must recall that the ages of the stars in our sample are subjected to important uncertainties, and that the reconstruction of their trajectories also becomes more uncertain as we trace them further into the past. Moreover, our estimates neglect streaming motions induced by the potential well associated with the spiral density wave and velocity jumps caused by large-scale spiral shocks that should reflect as deviations from the circular velocity \citep[e.g.,][]{Ramon18,Comeron97}. However, those estimates make the origin of at least the oldest of our red supergiants in the Sagittarius-Carina arm appear as a plausible hypothesis. In this scenario, the present massive star formation in Cygnus may be regarded as an indirect byproduct of star formation in the Sagittarius-Carina arm, when a group of massive stars formed there and, having drifted away from it as the density wave lagged behind the local galactic rotation, encountered the precursor of the present Cygnus giant molecular cloud complex. Although this is admittedly highly speculative, the apparent misfit between the Local Arm and the four main arms of our Galaxy might be explained if the Local Arm consisted of molecular material in the interarm region that survived as such the passage through the spiral density wave, and only became unstable against star formation when massive stars that had formed in a main arm entered it.

\section{Conclusions}

Our search for red supergiants has more than duplicated the number of such objects known in the direction of Cygnus, producing a homogeneously classified sample ranging from K2 to M4 spectral types. Of the 29 stars firmly established as supergiants in that interval, we assign 24 of them to the Cygnus region associated with the Local Arm. Four of the remaining five supergiants are most likely associated with the background Perseus arm, while the fifth is possibly a foreground star.

The presence of 24 red supergiants in the region, covering a wide range of masses, luminosities and ages, demonstrates that massive star formation in Cygnus started long before the OB associations that dominate the stellar component at present were born. We find no evidence of a spatial segregation between the younger and the older supergiants of our sample and, as expected from the internal velocity dispersion in OB associations, there is no obvious clustering among them and no particular proximity to the known OB associations.

Using the results of models of massive star evolution, we estimate the initial mass of each red supergiant based on its luminosity at present, and also the age based on the time at which red supergiants of various masses enter the helium-burning phase. These relationships contain uncertainties intrinsic to the models, and also due to their dependency on the initial rotation velocity, which is not known for our stars. With these caveats in mind, we have derived the star formation history of our sample. We find that the intense episode of star formation that has continued with the formation of the present-day OB associations started approximately 15~Myr ago, but we also find evidence for previous star formation episodes of less intensity having taken place between 20 and 30~Myr ago, and more than about 40~Myr ago. These episodes account for the existence of the less massive, cold and luminous red supergiants in the region. We find slightly different average kinematical properties between the younger and the older red supergiants: For an observer moving along the solar circle with the local standard of rest, the populations of young ($<20$~Myr) and old ($>20$~Myr) red supergiants would have appeared more separated in the past, with the older supergiants group moving on the average from lower to higher galactic longitudes, and the younger group moving in the opposite direction. We take this result as an indication that the red supergiants may not be simply tracing the star formation history of the Cygnus giant molecular complex. Instead, we favor an interpretation in which most of the older supergiants formed elsewhere outside the complex, entering it later from lower galactic longitudes. We propose that the oldest red supergiants currently observed in Cygnus actually started their lives as moderately massive stars in the Sagittarius-Carina main spiral arm of our Galaxy, drifting away from it as the density wave responsible for the spiral pattern made its progress across the stellar and gaseous disk. We present some crude estimates based on the parameters of the spiral structure of the Milky Way that support the plausibility of this scenario.

The exploitation of the Gaia astrometric legacy has just started, and the release of the final catalog of the mission in the coming years will make it possible to extend studies like the one presented here and investigate the properties of less massive members of Cygnus still on the main sequence, once nearly complete and nearly uncontaminated samples can be built. It may be expected that a much more detailed picture will emerge from such future studies, perhaps confirming our findings that hint to a longer and more complicated history of this complex than one may have suspected.

\begin{acknowledgements}

We are very pleased to thank the excellent support provided by the staff at the Calar Alto Observatory, especially on this occasion by Gilles Bergond, Ana Guijarro, and David Galad\'\i . The constructive review of the paper by the referee, Richard Boyle, is gratefully acknowledged. NS acknowledges support by the French ANR and the German DFG through the project "GENESIS" (ANR-16-CE92-0035-01/DFG1591/2-1). The Two Micron All Sky Survey (2MASS) is a joint project of the University of Massachusetts and the Infrared Processing and Analysis Center/California Institute of Technology, funded by the National Aeronautics and Space Administration and the National Science Foundation. This work has made use of data from the European Space Agency (ESA) mission
{\it Gaia} ({\tt https://www.cosmos.esa.int/gaia}), processed by the {\it Gaia} Data Processing and Analysis Consortium (DPAC, {\tt https://www.cosmos.esa.int/web/gaia/dpac/consortium}). Funding for the DPAC
has been provided by national institutions, in particular the institutions participating in the {\it Gaia} Multilateral Agreement. This research has made use of the SIMBAD database, operated at CDS, Strasbourg, France, and has made use of data products from the Midcourse Space Experiment. Processing of the data was funded by the Ballistic Missile Defense Organization with additional support from NASA Office of Space Science.

\end{acknowledgements}

\bibliographystyle{aa}

\bibliography{msg_cit}

\clearpage
\onecolumn
\longtab{3}{
\begin{landscape}
\begin{longtable}{lcccccccccl}
\caption{\label{supergiants}Cool supergiants in the direction of Cygnus}\\
\hline\hline
Star & type & $J$ & $K_S$ & $l$ & $b$ & $\tilde{\omega}$ & $\mu_\alpha \cos \delta$ & $\mu_\delta$ & $v_{\rm rad}$ & \\
     &      &     &       & $(^\circ)$ & $(^\circ)$ & (mas) & (mas~yr$^{-1}$) & (mas~yr$^{-1}$) & (km~s$^{-1}$) & \\
\noalign{\smallskip}\hline\noalign{\smallskip}
\endfirsthead
\caption{continued.}\\
\endhead
\endfoot
\noalign{\medskip {\centerline{Supergiants in the Cygnus associations}} \medskip}
J201842.17+411023.7 & K3   & $6.386 \pm 0.026$ & $3.787 \pm 0.322$ & $78.526$ & $ 2.937$ & $ 0.5186 \pm 0.1095$ & $-3.015 \pm 0.188$ & $-6.075 \pm 0.238$ & $ -25.4 \pm 0.33$ & \\
J201848.17+423637.4 & K7   & $ 4.91 \pm 0.278$ & $3.233 \pm 0.288$ & $79.728$ & $ 3.728$ & $ 0.4899 \pm 0.0364$ & $-3.278 \pm  0.06$ & $-4.644 \pm 0.061$ & $-14.91 \pm 0.44$ & 1 \\
J201857.52+390015.1 & K5   & $3.207 \pm 0.248$ & $ 1.92 \pm 0.246$ & $76.759$ & $ 1.676$ & $ 0.9516 \pm 0.0687$ & $-2.923 \pm 0.118$ & $-7.202 \pm 0.125$ & $ -6.06 \pm 0.21$ & 2 \\
J201905.19+394715.9 & K7   & $5.122 \pm 0.037$ & $3.448 \pm 0.268$ & $77.421$ & $ 2.097$ & $ 0.6138 \pm 0.1246$ & $-2.031 \pm 0.249$ & $-3.916 \pm 0.225$ & $-18.28 \pm 0.45$ &  \\
J201917.19+382550.2 & K2   & $5.151 \pm 0.037$ & $ 3.94 \pm 0.029$ & $76.321$ & $   1.3$ & $ 0.4892 \pm 0.0333$ & $ -3.52 \pm 0.059$ & $-5.124 \pm 0.071$ & $ -9.81 \pm 0.19$ & 3 \\
J201919.97+341101.0 & K2   & $5.105 \pm 0.274$ & $3.903 \pm 0.272$ & $72.821$ & $-1.105$ & $ 0.8844 \pm 0.0315$ & $ 2.653 \pm 0.046$ & $ 0.021 \pm 0.058$ & $-10.48 \pm 0.19$ & 4 \\
J202114.07+353716.5 & M2   & $3.283 \pm 0.234$ & $1.623 \pm 0.202$ & $74.227$ & $-0.612$ & $ 0.6054 \pm 0.0671$ & $-2.846 \pm 0.109$ & $-4.499 \pm 0.118$ & $-11.56 \pm 0.47$ & 5 \\
J202121.92+365555.5 & M4   & $2.345 \pm 0.256$ & $0.724 \pm 9.996$ & $ 75.32$ & $ 0.112$ & $ 0.7269 \pm 0.0811$ & $-2.929 \pm  0.14$ & $-5.223 \pm 0.151$ & $ -6.82 \pm 0.42$ & 6 \\
J202133.14+363654.4 & K2   & $5.924 \pm 0.018$ & $ 3.88 \pm 0.364$ & $ 75.08$ & $-0.099$ & $ 0.4769 \pm 0.1247$ & $-3.069 \pm 0.214$ & $-5.056 \pm 0.237$ & $ -16.9 \pm 0.34$ & \\
J202138.55+373158.9 & M4   & $2.117 \pm 0.276$ & $0.299 \pm 0.206$ & $75.845$ & $ 0.409$ & $ 0.7517 \pm 0.1009$ & $-3.856 \pm 0.155$ & $-5.835 \pm 0.206$ & $-22.85 \pm 0.45$ & 7 \\
J202211.42+405121.2 & K2   & $5.075 \pm 0.018$ & $3.327 \pm 0.354$ & $78.637$ & $ 2.215$ & $ 0.5337 \pm 0.0494$ & $-2.623 \pm 0.088$ & $-4.134 \pm 0.088$ & $-20.76 \pm 0.18$ & \\
J202308.60+365145.0 & M4   & $4.955 \pm 0.262$ & $2.606 \pm 0.338$ & $75.465$ & $ -0.22$ & $ 0.4548 \pm 0.1096$ & $-3.295 \pm 0.187$ & $-6.597 \pm 0.205$ & $  -9.3 \pm 0.63$ & 8 \\
J202502.94+433605.9 & M2   & $5.603 \pm 0.021$ & $3.851 \pm 0.286$ & $81.194$ & $ 3.353$ & $ 0.4829 \pm 0.0919$ & $-2.072 \pm 0.165$ & $-5.258 \pm 0.153$ & $ -4.75 \pm 0.28$ & \\
J202558.05+382107.6 & M4   & $2.421 \pm 0.268$ & $ 0.42 \pm 0.224$ & $77.005$ & $ 0.181$ & $ 0.9151 \pm  0.092$ & $-3.574 \pm 0.18 $ & $-6.279 \pm 0.159$ &                   & 9 \\
J202603.31+410827.6 & M1.5 & $4.566 \pm 0.244$ & $2.415 \pm 0.314$ & $ 79.29$ & $ 1.781$ & $ 0.5799 \pm 0.1036$ & $-2.679 \pm 0.167$ & $ -3.94 \pm 0.178$ & $-11.23 \pm 0.64$ & 10 \\
J202645.17+381343.5 & K5   & $4.952 \pm  0.25$ & $3.243 \pm 0.384$ & $76.994$ & $-0.016$ & $ 0.8296 \pm  0.048$ & $-0.421 \pm 0.077$ & $-3.099 \pm 0.074$ & $-11.76 \pm 0.14$ & \\
J202736.78+385046.1 & K7   & $4.682 \pm 0.206$ & $3.083 \pm 0.302$ & $77.594$ & $ 0.206$ & $ 0.6128 \pm 0.0417$ & $ 0.005 \pm 0.068$ & $-2.819 \pm  0.07$ & $ -11.7 \pm 0.15$ & 11 \\
J203301.05+404540.4 & M2   & $3.859 \pm  0.29$ & $1.296 \pm 0.276$ & $79.753$ & $ 0.494$ & $ 1.0224 \pm 0.1339$ & $-1.877 \pm 0.233$ & $-4.245 \pm 0.261$ & $ -1.18 \pm 2.16$ & 12 \\
J203323.91+403644.2 & K7   & $6.955 \pm 0.024$ & $3.997 \pm 0.015$ & $79.677$ & $ 0.347$ & $-0.2561 \pm 0.1144$ & $-1.245 \pm 0.191$ & $-4.235 \pm 0.186$ & $-21.06 \pm 1.41$ & \\
J203401.43+422526.6 & M3.5 & $3.942 \pm  0.24$ & $1.013 \pm 0.226$ & $81.201$ & $ 1.333$ & $ 1.0667 \pm 0.1532$ & $-1.064 \pm 0.301$ & $-5.265 \pm 0.299$ & $-19.21 \pm 1.05$ & 13 \\
J203936.52+391209.7 & M4   & $4.697 \pm 0.244$ & $1.867 \pm 0.296$ & $79.269$ & $-1.452$ & $ 1.0329 \pm 0.1236$ & $-1.651 \pm 0.217$ & $-4.859 \pm 0.217$ & $-20.47 \pm 0.42$ & 14 \\
J204340.64+442837.6 & M2   & $4.808 \pm  0.32$ & $2.449 \pm  0.29$ & $83.897$ & $  1.19$ & $ 0.3994 \pm 0.0939$ & $-2.546 \pm 0.172$ & $ -5.18 \pm 0.181$ & $-17.13 \pm 1.36$ & 15 \\
J204604.55+445209.4 & K7   & $3.954 \pm 0.208$ & $2.448 \pm  0.27$ & $84.469$ & $ 1.099$ & $ 0.5864 \pm 0.0553$ & $-1.896 \pm 0.084$ & $-3.298 \pm 0.095$ & $-30.34 \pm 0.77$ & 16 \\
J205425.72+431247.0 & M0   & $5.418 \pm 0.024$ & $3.787 \pm 0.308$ & $84.141$ & $-1.112$ & $ 0.4853 \pm 0.0805$ & $ -2.25 \pm 0.138$ & $-3.226 \pm 0.146$ & $-14.77 \pm 0.16$ & \\
\hline
\noalign{\medskip {\centerline{Suspected foreground and background supergiants}} \medskip}

Star & type & $J$ & $K_S$ & $l$ & $b$ & $\tilde{\omega}$ & $\mu_\alpha \cos \delta$ & $\mu_\delta$ & $v_{\rm rad}$ \\
     &      &     &       & $(^\circ)$ & $(^\circ)$ & (mas) & (mas~yr$^{-1}$) & (mas~yr$^{-1}$) & (km~s$^{-1}$) \\
\noalign{\medskip}\hline\noalign{\medskip}
J201126.21+354227.7 & M4   & $4.463 \pm 0.306$ & $2.602 \pm 0.306$ & $73.187$ & $ 1.087$ & $0.2385 \pm 0.0779$ & $-3.075 \pm 0.133$ & $-5.256 \pm 0.146$ & $               $ & 17 \\
J202412.43+411402.9 & M4   & $5.365 \pm 0.286$ & $3.246 \pm 0.346$ & $79.165$ & $ 2.119$ & $0.8543 \pm 0.0869$ & $-1.748 \pm 0.155$ & $-0.599 \pm 0.160$ & $-44.79 \pm 0.39$ & \\

J203817.18+420425.1 & K2   & $4.538 \pm 0.254$ & $3.252 \pm 0.322$ & $81.396$ & $ 0.492$ & $1.1699 \pm 0.0357$ & $ 0.347 \pm 0.062$ & $-1.162 \pm 0.061$ & $ -5.37 \pm 0.16$ & 18 \\
J204636.65+424421.6 & M2   & $6.029 \pm 0.018$ & $ 3.99 \pm 0.036$ & $82.867$ & $-0.308$ & $0.3744 \pm 0.1168$ & $-2.669 \pm 0.181$ & $-4.163 \pm 0.198$ & $-58.66 \pm 0.44$ & \\
J204924.95+443349.1 & M3.5 & $7.076 \pm 0.021$ & $3.987 \pm  0.34$ & $84.604$ & $ 0.446$ & $0.2337 \pm 0.1871$ & $-2.675 \pm 0.327$ & $-4.084 \pm 0.311$ & $-60.98 \pm 2.59$ & 19 \\
\noalign{\smallskip}\hline\noalign{\smallskip}
\end{longtable}
\begin{tablenotes}
\item 1: BD$+42^\circ$~3864
\item 2: HD~193469, classified as K5Ib \citep{Bidelman57,Griffin60}
\item 3: HD~228898, classified as K5 in the Henry Draper catalog
\item 4: HD~228890, classified as K8 in the Henry Draper catalog
\item 5: V1749~Cyg, classified as M2.5I in \citet{Levesque05}
\item 6: BI~Cyg, classified as M4Iab in \citet{Gahm72}
\item 7: BC~Cyg, classified as M3I \citep{Levesque05}, M4Ia \citep{Gahm72}
\item 8: Classified as M5I in \citet{Grasdalen79}
\item 9: KY~Cyg, classified as M3I-M4I in \citep{Levesque05}
\item 10: Classified as M2 in \citet{Nassau54}
\item 11: V1388~Cyg
\item 12: Haro-Chavira~2, classified as M4I in \citet{Grasdalen79}
\item 13: Haro-Chavira~1, classified as M1I in \citet{Grasdalen79}
\item 14: Classified as M4I in \citet{Grasdalen79} \item 15: V2429~Cyg, classified as M3I in \citet{Grasdalen79}
\item 16: RR~Cyg, classified as M1 in \citet{Cameron56}
\item 17: V430~Cyg, classified as M4 in \citet{Nassau54}
\item 18: HD~196819, classified as K2.5IIb in \citet{Keenan89}
\item 19: Listed as carbon star candidate in \citet{Stephenson89,Chen97} \end{tablenotes}
\end{landscape}
}

\end{document}